\documentclass[a4paper,12pt]{article}

\usepackage{geometry,slashed}
\usepackage{amssymb}

\RequirePackage[T1]{fontenc} 

\RequirePackage{graphicx}
\RequirePackage{flushend}
\RequirePackage[numbers,sort&compress]{natbib}
\RequirePackage[colorlinks,citecolor=blue,urlcolor=blue,linkcolor=blue]{hyperref}

\usepackage{yfonts}
\usepackage{booktabs}
  \usepackage{soul}
  \usepackage[usenames,dvipsnames]{xcolor}
 
 \definecolor{X575}{rgb}{0.05, 0.7, 0.05}
 
\usepackage{cancel} 
\usepackage{graphicx} 
 \usepackage{pgffor}
\usepackage{array}
\usepackage{amsmath}
 \usepackage{color}
 \usepackage{hyperref}
 \usepackage{authblk}
 \usepackage{epstopdf}
 \usepackage{enumerate}
 \usepackage{multirow}
 \usepackage{siunitx}
 \usepackage{float}

  \usepackage{listings}
  \usepackage{calrsfs}
  \usepackage{subfigure}

 \newcommand{\tllj}{t\ell^+\ell^-j}
 
 \newcommand{\mll}{m(\ell^+\ell^-)}
 \newcommand{\MSbar}{{\rm \overline{MS}}}
  \newcommand{\TO}{\rightarrow}

\newcommand{\gev}{\textrm{GeV}}

\newcommand{\mb}{m_b}

\newcommand{\mz}{m_Z}

\newcommand{\gamh}{\Gamma_H}
\newcommand{\gmu}{G_\mu}

\newcommand\mf{{\sc\small MadFKS}}
\newcommand\ml{{\sc\small MadLoop}}
\newcommand\ct{{\sc\small CutTools}}
\newcommand\nin{{\sc\small Ninja}}
\newcommand\coll{{\sc\small Collier}}

\def\alphas{\alpha_s}

\def\TO{{\longrightarrow}}
\def\ord{\mathcal{O}}
\def\alphas{\alpha_s}
\def\LO{\rm LO}
\def\NLO{\rm NLO}
\def\NLOQCD{\rm NLO_{QCD}}
\def\NLOQCDt{\NLO_{{\rm QCD},t{\rm -ch.}}}
\def\NLOQCDEW{\NLO_{{\rm QCD+EW}}}
\def\NLOEW{\rm NLO_{EW}}

\def\PSQCD{\rm PS_{QCD}}
\def\PSQCDQED{\rm PS_{{QCD+QED}}}

\title{NLO QCD+EW predictions for  $tHj$ and $tZj$ production at the LHC}

\author[1]{Davide Pagani\thanks{davide.pagani@desy.de}}
\author[2]{Ioannis Tsinikos\thanks{ioannis.tsinikos@thep.lu.se}}
\author[3]{Eleni Vryonidou\thanks{eleni.vryonidou@cern.ch}}

\affil[1]{\small DESY, Theory Group, Notkestrasse 85, 22607 Hamburg, Germany}
\affil[2]{\small Theoretical Particle Physics, Department of Astronomy and Theoretical Physics, Lund University, S\"olvegatan 14A, SE-223 62 Lund, Sweden}
\affil[3]{\small Theoretical Physics Department, CERN, CH-1211 Geneva 23, Switzerland}

\begin{document}
 
  \date{}
 
 \maketitle
 
  \vspace*{-9cm}
  {
      {{\color{blue}{DESY 20-101}  \hfill
      \color{blue}{LU-TP 20-33}
      \hfill \color{blue}{CERN-TH-2020-090}}
  }
  \vspace*{9cm}

\begin{abstract}
In this work we calculate the cross sections for the hadroproduction of a single top quark or antiquark in association with a Higgs ($tHj$) or $Z$ boson ($tZj$) at NLO QCD+EW accuracy. In the case of $tZj$ production we consider both the case of the $Z$ boson undecayed and the complete final state $t \ell^+ \ell^- j$, including off-shell and non-resonant effects. We perform our calculation in the five-flavour-scheme (5FS), without selecting any specific production channel ($s$-, $t$- or $tW$ associated). Moreover, we provide a more realistic estimate of the theory uncertainty by carefully including the differences between the  four-flavour-scheme (4FS) and 5FS predictions. The difficulties underlying this procedure in the presence of EW corrections are discussed in detail.  We find that NLO EW corrections are in general within the NLO QCD theory uncertainties only if the flavour scheme uncertainty (4FS vs.~5FS) is taken into account. For the case of $t \ell^+ \ell^- j$ production we also investigate differences between NLO QCD+EW predictions and NLO QCD predictions matched with a parton shower simulation including multiple photon emissions.  \end{abstract}

 \section{Introduction}\label{sec:intro}

The study of the interactions of the top quark, gauge and Higgs bosons is one of the main goals of TeV-scale colliders. After the discovery of the Higgs boson in the Run I, the Large Hadron Collider (LHC) has 
reached a golden era of precision physics and will continue to stress-test the predictions of the  Standard
 Model (SM) of elementary particles.  The success of this ambitious research program critically depends on a collaborative theoretical and experimental effort to detect deviations and/or constrain new physics, 
 with sensitivities that reach the multi-TeV scale. 
The LHC has opened new possibilities, allowing for the first time the measurement of processes that directly probe  the interaction of the 
top quark with both the neutral gauge bosons and the Higgs boson. This set of processes consists of the associated production of a single top or a 
top-quark pair with a Higgs or a neutral gauge boson. Since these processes probe some of the least known interactions in the SM, they have attracted 
considerable attention both on the experimental and theoretical side.
The leading production mechanism for top-pair associated production is through QCD interactions (at order $\alpha_s^2 \alpha$ at 
the Born level). Single top associated production, which is the focus of this work, proceeds instead via electroweak interactions. Whilst single 
top production ($tj$) has only a rate of roughly one third of the strong $t \bar t$ production, $tZj$  has a cross section similar to the 
$t\bar{t}Z$ one. Indeed, single top production in association with a $Z$ boson has already been measured at the LHC by both CMS 
\cite{Sirunyan:2018zgs, Sirunyan:2017nbr} and ATLAS \cite{ Aaboud:2017ylb, Aad:2020wog}. Finally the $tHj$ rate is about 
10\% of the $t\bar{t}H$ one, and the process has been searched for at the LHC \cite{Khachatryan:2015ota, CMS-PAS-HIG-16-019, Sirunyan:2018lzm}.

The fact that this set of processes plays an important role in the search for new top-quark interactions has been studied extensively
 by the theory community~\cite{Maltoni:2001hu, Biswas:2012bd, Farina:2012xp, Demartin:2015uha}. They open up a unique 
 possibility of probing top-Higgs, top-gauge, triple-gauge, gauge-Higgs interactions in the same final state. The complete analysis of 
 the $tZj$ and $tHj$ processes in the Standard Model Effective Field Theory (SMEFT) framework, including next-to-leading (NLO) QCD corrections,
 was presented in Ref.~\cite{Degrande:2018fog}. This work demonstrated the importance of single top associated production as a probe 
 of new interactions, and showed the crucial role these processes can play in a global SMEFT interpretation of top
 couplings.  Another important finding was that differential distributions can be particularly sensitive to modifications of the top-quark interactions, 
 therefore differential measurements of these processes can prove highly beneficial in constraining new physics effects. 

The special role of single top associated production as a probe of new interactions as well as the increasing precision of 
the experimental inclusive measurements and the prospect of differential measurements demand a precise theoretical 
description of these processes within the Standard Model.  This class of processes are ``purely'' electroweak, with the  consequence
that the QCD corrections are typically small and under control. The SM predictions at NLO in QCD for $tZj$ and $tHj$ were first presented in 
Refs.~\cite{Campbell:2013yla} and~\cite{Demartin:2015uha}, respectively. Whilst Ref.~\cite{Demartin:2015uha} performs a detailed 
comparison between the four flavour scheme (4FS) and five flavour scheme (5FS) computations for $tHj$ at both the inclusive and differential level, a corresponding exploration of the $tZj$ process is lacking. 

Motivated by the increasing precision of experimental measurements, the main goal of this paper is to revisit single top 
associated production within the SM, computing for the first time NLO QCD+EW corrections in the 5FS and examining the impact of higher-order corrections at both the inclusive and differential level for $tHj$, $tZj$ and more in general $\tllj$ production. As a matter of fact, the results of our calculation correspond to one of the items of the 2019 Les Houches precision Standard Model wish list \cite{Amoroso:2020lgh}. Moreover, similarly to Ref.~\cite{Demartin:2015uha}, we perform a detailed 
comparison between the 4FS and 5FS computations and we design a strategy for taking into account their differences as an additional uncertainty, together with scale uncertainties and EW corrections. In the case of $\tllj$ production, where two leptons are present in the final state, we also compare NLO QCD+EW predictions at fixed order with NLO QCD predictions matched with parton shower effects including QED photon emissions from fermions.

This paper is organised as follows. In Sec.~\ref{sec:setup} we describe in detail our calculation setup. 
First, in Sec.~\ref{sec:flavscheme}, we discuss the differences between the 4FS and 5FS and we explain the strategy we have designed for taking into account their differences. In Sec.~\ref{sec:ch-sepa} we discuss the technical difficulties for separating $t$-, $s$- and $tW$ associated modes when EW corrections are calculated, which partly motivate the aforementioned strategy. Input parameters are specified in Sec.~\ref{sec:input}. In Sec.~\ref{sec:results} we provide and comment on numerical results obtained within {\sc\small MadGraph5\_\-aMC@NLO} \cite{Alwall:2014hca, Frederix:2018nkq} and corresponding to the most precise predictions for $tHj$, $tZj$ and $\tllj$ production to-date. In Sec.~\ref{sec:inclusive} we provide the predictions for total cross sections at NLO QCD+EW accuracy, while differential distributions are analysed in Sec.~\ref{sec:differential}. In Sec.~\ref{sec:shower} we focus on the $\tllj$ process and discuss the differences between the fixed-order result at NLO QCD+EW accuracy and the NLO QCD prediction matched with a parton shower taking into account multiple photon emissions from fermions. We give our conclusions and outlook in Sec.~\ref{sec:conclusion}.

 \section{Calculation Setup}
 \label{sec:setup}

In this section we describe the calculation setup, which is common for the three processes considered in this work:
\begin{itemize}
\item $pp~\TO ~tHj+\bar tHj$,
\item $pp~\TO~ tZj+\bar tZj$,
\item $pp~\TO ~\tllj + \bar t\ell^+\ell^- j$.
\end{itemize} 
In the following, unless it is differently specified, with the notation $tHj$, $tZj$ and $\tllj$ we will understand both the final states with top quarks and antiquarks.    
We remind the reader that the production process $\tllj$ corresponds to the production process $tZj$ only in the limit $\mll\TO\mz$, {\it i.e.}, in the narrow-width approximation. In general, off-shell and interference effects are present, especially, the $\ell^+\ell^-$ pair can also stem from an off-shell photon.

We calculate cross sections, at the inclusive and differential level, at NLO QCD+EW accuracy in the 5FS.
 Expanding in powers of $\alpha_s$ and $\alpha$ and
following the notation already used in Refs.~\cite{Frixione:2014qaa, Frixione:2015zaa, Pagani:2016caq, Frederix:2016ost, Czakon:2017wor, Frederix:2017wme, Frederix:2018nkq, Broggio:2019ewu, Frederix:2019ubd}, the different contributions to any differential or inclusive cross section $\Sigma$ can be denoted as:
\begin{align}
\Sigma^{}_{\LO}(\alpha_s,\alpha) &=  \alpha^{3+k} \Sigma_{3+k,0}^{} \nonumber\\
 &\equiv \LO_1\, , \label{eq:blobLO} \\
 \Sigma^{}_{\NLO}(\alpha_s,\alpha) &= \alphas \alpha^{3+k} \Sigma_{4+k,0}^{}+ \alpha^{4+k} \Sigma_{4+k,1}^{} \nonumber\\
 &\equiv \NLO_1 + \NLO_2 \, , \label{eq:blobNLO} 
\end{align}
where $k=0$ for $tHj$ and $tZj$ and $k=1$ for $\tllj$. 
For convenience,
we will also use the standard notation $\NLOQCD$ and $\NLOQCDEW$ for denoting the quantities  $\LO_1+\NLO_1$ and $\LO_1+\NLO_1+\NLO_2$, respectively. In other words, the $\NLO_1$ and $\NLO_2$ terms are the NLO QCD and NLO EW corrections, respectively.  One should note that no additional perturbative orders are present when all possible SM tree-level and one-loop diagrams contributing to these processes are taken into account. In other words, the complete-NLO, {\it i.e.}  the set of all the possible contributions of $\ord(\alpha_s^n \alpha_s^m)$ with $m,n>0$ and $m+n\leq4+k$, and $\NLOQCDEW$ predictions coincide. Similarly, the $\LO_1$ contribution coincides with the $\LO$ prediction.\footnote{These statements would not hold true  in the 4FS, unless  photon-initiated processes are artificially neglected. Indeed, any $b$-initiated contribution of order $\alpha_s^n \alpha^m$ in the 5FS is related to a corresponding  gluon-initiated contribution of order $\alpha_s^{n+1} \alpha^m$ in the 4FS, via the $g\TO b \bar b$ splitting. However, it is also related to a corresponding  photon-initiated contribution of order $\alpha_s^n \alpha^{n+1}$, via the $\gamma\TO b \bar b$ splitting. Therefore, following the notation already used in Refs.~\cite{Frixione:2014qaa, Frixione:2015zaa, Pagani:2016caq, Frederix:2016ost, Czakon:2017wor, Frederix:2017wme, Frederix:2018nkq, Broggio:2019ewu, Frederix:2019ubd}, also an $\LO_2$ and  an $\NLO_3$ term would be present.} We note here that this is not the case for other relevant processes in top-quark physics, such as, {\it e.g.}, $t \bar t W$ and $t \bar t t \bar t$ for which the two approximations are different and lead to sizeable numerical differences due to contributions that are formally suppressed w.r.t.~the $\NLOEW$ in the $(\alpha_s/\alpha)$ power counting \cite{Frederix:2017wme,  Broggio:2019ewu, Kulesza:2020nfh, Frederix:2020jzp}.

 \subsection{Flavour-scheme and scale uncertainties}
 \label{sec:flavscheme}

As can be seen in Eq.~\eqref{eq:blobLO}, LO predictions do not depend on $\alpha_s$; for purely electroweak processes the dependence on $\alpha_s$ enters only via the NLO QCD corrections, the $\NLO_1$ contribution in Eq.~\eqref{eq:blobNLO}, or higher-order terms in the perturbative expansion. For this reason, while the explicit dependence on the factorisation scale $\mu_F$ is present already at LO, the explicit dependence on the renormalisation scale $\mu_R$ enters only at NLO QCD accuracy. Therefore, the naive approach of scale variations would in turn lead to artificially small QCD scale uncertainties both at LO and NLO accuracy. 
This fact is very important also because the NLO EW corrections, the $\NLO_2$, can be potentially larger than the residual scale uncertainties that are obtained via the independent variation of $\mu_F$ and $\mu_R$ by a factor of two.

A calculation at next-to-next-to-leading-order (NNLO) QCD accuracy would in principle help, but at the moment is unfeasible. Indeed, no NNLO QCD calculation for a $2\TO3$ process with a massive coloured particle in the final state has ever been performed. In order to amend this situation, we follow the approach of Ref.~\cite{Demartin:2015uha}, where the case of $tHj$ production has  been considered. This approach relies on the fact that predictions at LO and especially NLO QCD accuracy can be calculated in the 5FS, but also in the 4FS,  and the difference between the two predictions at NLO QCD accuracy can be interpreted as an additional estimate of the uncertainty due to missing higher-order terms in $\alpha_s$. We briefly recall in the following the rationale behind this interpretation.

Given any scale $Q$ involved in the process, besides power corrections of order $(\mb/Q)^n$ with $n>0$, which can be taken into account only in the 4FS, the two flavour schemes effectively correspond to a rearrangement of logarithms of the form $\alpha_s^n\log^k(Q/\mb)$ with at most $k\leq n$.  For $Q\gg\mb$, {\it i.e.} for instance for total cross sections without hard cuts such as jet vetoes or distributions far from the threshold region, power corrections are negligible and the aforementioned logarithms are large. In the 5FS, these large logarithms are resummed and therefore better estimated. On the other hand, at variance with the 5FS, in the 4FS the $\mu_R$ dependence enters already at LO for the processes we are considering and therefore NLO calculations involve a truly NLO dependence on  $\mu_R$. Moreover, at the differential level, the presence of an additional particle in the final state resembles a more realistic signature and, {\it e.g.}, eliminates or at least softens some of the sharp edges or endpoints that are typical of fixed-order calculations, where the recoil momentum of any particle is shared among the few others in the final state. Last but not least, close to the threshold, power corrections are taken into account. 
See also Refs.~\cite{Maltoni:2012pa,Lim:2016wjo} for a more detailed and general discussion on the differences between 4FS and 5FS prediction for $b$-initiated processes.

As already said, in order to achieve a more realistic estimate of the uncertainty due to missing higher-order terms in $\alpha_s$, we will take into account the flavour-scheme dependence in our theory uncertainty, as done in Ref.~\cite{Demartin:2015uha}. However, we cannot straightforwardly extend this strategy to our calculation. Indeed, at variance with the study presented here, Ref.~\cite{Demartin:2015uha} focussed only on the $t$-channel mode in $tHj$ production and especially did not take into account NLO EW corrections, the $\NLO_2$ contribution in Eq.~\eqref{eq:blobNLO}.  Therefore, we need to slightly modify this strategy in order to adapt it to our set-up. In the following, we describe how we calculate and combine theory uncertainties related to  the flavour-scheme dependence and scale variations. Afterwards, in the next section, we discuss the problems related to the separation of the different production modes ($t$-channel, $s$-channel and $W$ boson associated production) and we motivate on the basis of those problems the rationale behind the strategy that we have adopted.

\begin{figure}[!t]
    \centering
    \subfigure[$t$-channel]
    {
        \includegraphics[ height=3.5cm]{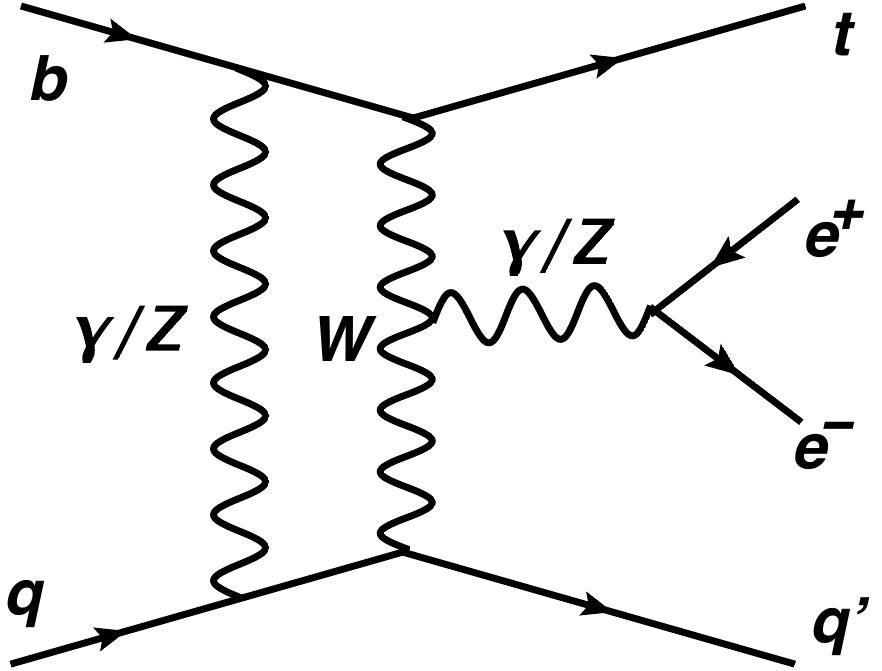}
        \label{fig:tch}
    }
    \hspace{1cm}
    \subfigure[$s$-channel]
    {
        \includegraphics[height=3.5cm]{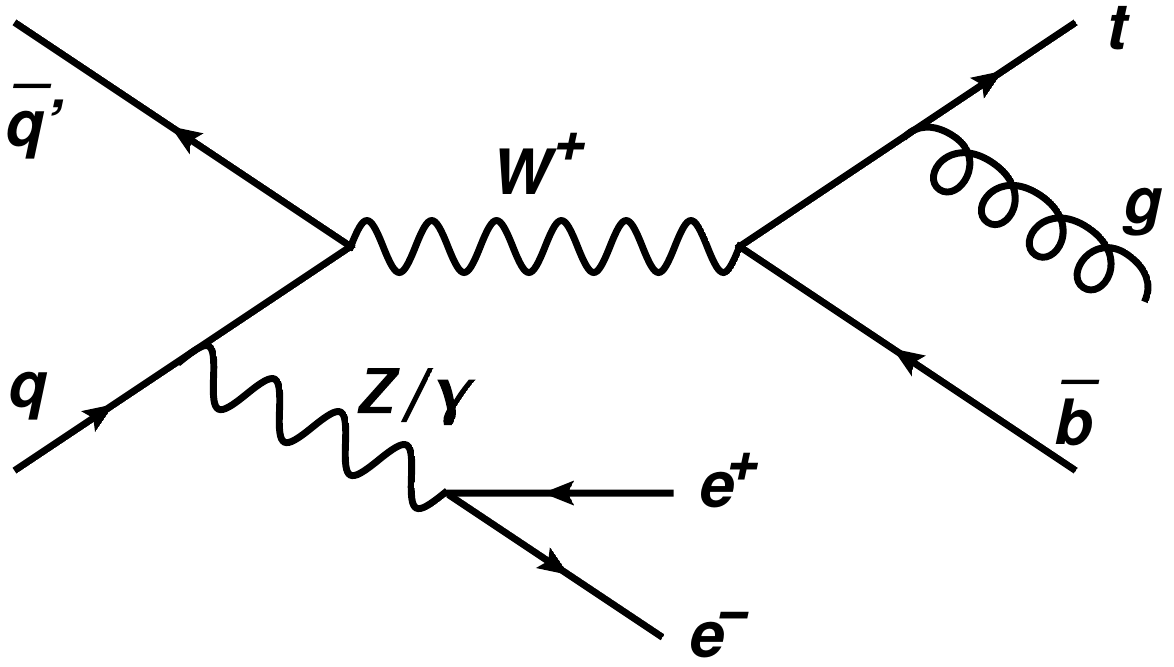}
        \label{fig:sch}
    }
    \subfigure[$tW$]
    {
        \includegraphics[ height=3.5cm]{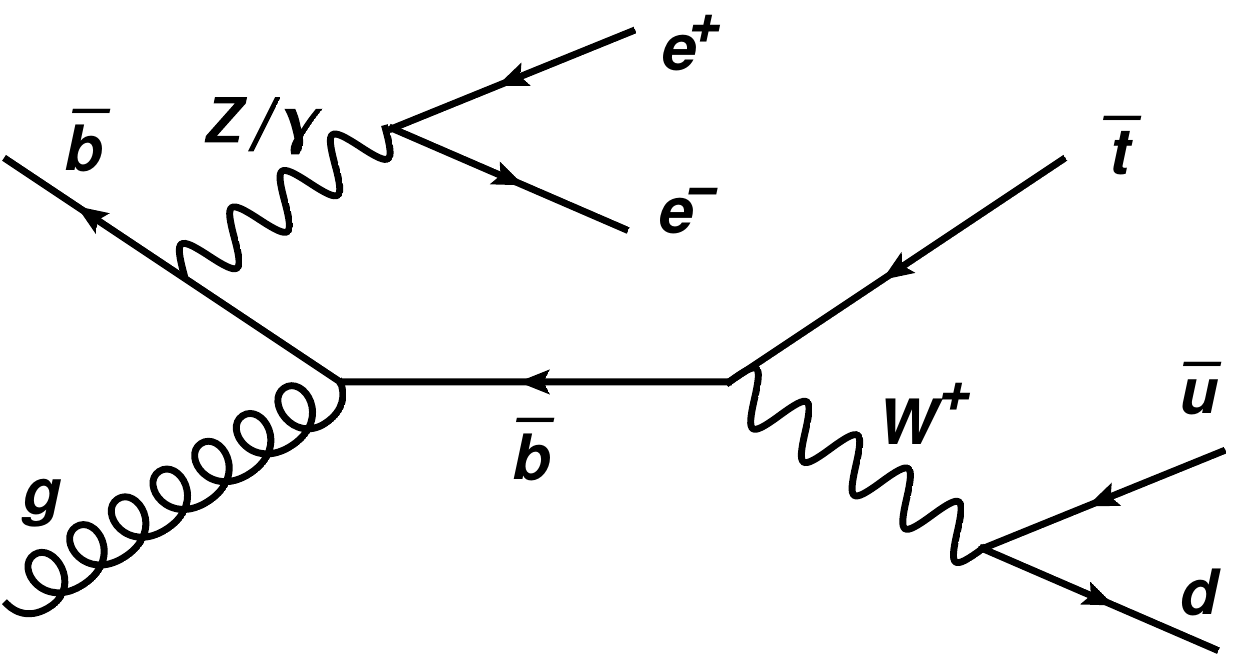}
        \label{fig:tW}
    }
    \caption{Representative Feynman diagrams for the different channels entering the $\NLOQCDEW$ predictions for $\tllj$ production. The diagram \ref{fig:tch} contributes to the $\NLO_2$, while the diagrams \ref{fig:sch} and \ref{fig:tW} contribute to the $\NLO_1$. Similar diagrams are present for $tZj$ production, while in $tHj$ production the Higgs boson does not couple to the initial-state particles. }
    \label{fig:sample_subfigures}
\end{figure}

For any observable, we define as $\NLOQCDt^{4FS}$ and $\NLOQCDt^{5FS}$ the prediction at NLO QCD accuracy where only the $t$-channel diagrams have been taken into account. There is no difference for the $tHj$, $tZj$ and $\tllj$ production, so we do not specify them in the following discussion. With the notation $(\NLOQCDt^{4FS})^{+\Delta_+^{\rm 4FS}}_{-\Delta_-^{4FS}}$ and $(\NLOQCDt^{5FS})^{+\Delta_+^{\rm 5FS}}_{-\Delta_-^{5FS}}$ we indicate, together with the central value,  the {\it relative} difference with the upper ($\Delta_+$) and lower ($\Delta_-$) values of the scale-uncertainty band, which we evaluate via the 9-point independent variation of the renormalisation and factorisation scale. Similarly, $(\NLOQCDEW^{5FS})^{+\tilde{\Delta}_+^{\rm 5FS}}_{-\tilde{\Delta}_-^{5FS}}$ denotes with a similar notation the prediction, with scale uncertainties, at NLO QCD+EW accuracy in the 5FS , where the tilde on top of $\Delta$ has been added just for distinguishing them from the purely QCD case. We notice that  $(\NLOQCDEW^{5FS})^{+\tilde{\Delta}_+^{\rm 5FS}}_{-\tilde{\Delta}_-^{5FS}}$ is not obtained by selecting $t$-channel diagrams, but retaining all the possible contributions: not only $t$-channel, but also $s$-channel and $tW$ associated production with subsequent $W$ boson hadronic decay, see Fig.~\ref{fig:sample_subfigures} for representative diagrams.

In order to combine scale and flavour-scheme uncertainties at NLO QCD accuracy, we consider the $t$-channel only and we define the quantity $(\NLOQCDt^{5FS})^{+\Delta_+^{\rm 4-5FS}}_{-\Delta_-^{4-5FS}}$  via the envelope of the two bands given by $(\NLOQCDt^{4FS})^{+\Delta_+^{\rm 4FS}}_{-\Delta_-^{4FS}}$ and $(\NLOQCDt^{5FS})^{+\Delta_+^{\rm 5FS}}_{-\Delta_-^{5FS}}$, where the central value is set equal to the one in the 5FS. 
The quantities $\Delta_+^{\rm 4-5FS}$ and $\Delta_-^{\rm 4-5FS}$ are then propagated to the $\NLOQCDEW$ prediction in the 5FS.
In conclusion, in order to combine flavour-scheme and scale uncertainties and take into account EW corrections, not only for $t$-channel, we will employ as reference prediction the quantity
$(\NLOQCDEW^{5FS})^{+\Delta^{\rm 4-5FS}_+}_{-\Delta^{\rm 4-5FS}_-}$
 and in the case of QCD only corrections, in order to be consistent, we will use the quantity  $(\NLOQCD^{5FS})^{+\Delta^{\rm 4-5FS}}_{-\Delta^{4-5FS}}$, where in the quantity $\NLOQCD^{5FS}$ the requirement of $t$-channel only is not applied. In Sec.~\ref{sec:results} predictions obtained following this approach will be simply denoted by  5FS$_{\rm 4-5}^{\rm scale}$.

\subsection{Separation of different production modes}

\label{sec:ch-sepa}

In this section we explain why we cannot select the $t$-channel mode and  at the same time take into account NLO EW corrections. Moreover, we  explain why we believe that not singling out the $t$-channel mode is anyway preferable for providing reference predictions for experimental measurements. 
After this explanation, we will motivate the strategy that we have designed in order to take into account flavour-scheme dependence and scale variations in our theory uncertainty.

First of all, it is important to point out that the division of single top production into $t$-channel, $s$-channel and $tW$ associated production is conventional and especially it is clearly defined only under certain conditions that depend on the flavour-scheme choice and the perturbative order of the calculation. The presence of a Higgs boson or a $Z$ boson or an $\ell^+\ell^-$ pair is irrelevant for the present discussion and therefore it is understood in the following.

 In the 5FS, at LO, the production of a single top quark can be divided into the three categories according to the $W$ boson line in the Feynman diagrams: $s$-channel propagator, $t$-channel propagator or final-state associated production. This division is gauge invariant and well defined also at NLO QCD accuracy, however the $tW$ associated production leads to the same final state as $t \bar t$ production, which has a much larger cross section, and interferes with it. This occurs when $b$-quark real emission induces diagrams with an additional top decaying into a $Wb$ pair. At NNLO QCD accuracy, even $s$- and $t$-channel topologies start to interfere. When EW corrections are taken into account, the situation becomes more complex. The main reason is that photons in the initial state have to be taken into account. The process $\gamma q\TO t \bar b q'$, contributing to the $\NLO_2$ term in Eq.~\eqref{eq:blobNLO}, involves $t$-channel diagrams, $s$-channel diagrams and the separation is not gauge invariant. Similarly, the process $\gamma b \TO t q\bar q '$,  contributing also to the $\NLO_2$ term, involves $t$-channel diagrams and diagrams where a $q\bar q '$ pair stems from an $s$-channel $W$ boson, and also in this case the separation is not gauge invariant. In other words, similarly to the case of $tW$ at NLO QCD, where $t \bar t$ diagrams emerge in real-emission corrections, in the NLO EW corrections to $tj$ production in the 5FS we cannot separate in a straightforward and gauge-invariant way the contribution from different production modes.  
In the 4FS, the situation is even worse. Even with only QCD corrections, the $t$-channel at NLO can interfere with the $s$-channel at NNLO and with the $W$-associated production with $W\TO q\bar q '$ decays. Moreover, the $tW$ and $t \bar t$ interference is already present at the tree level.

From the previous discussion it is manifest why singling out contributions from $t$-channel production is very challenging in our calculation. Moreover, we believe that independently of these difficulties, singling out $t$-channel production should not be done when providing reference values for experimental measurements such as those already performed by the ATLAS and CMS collaboration. Indeed, in these analyses,  selection cuts have not been designed  to separate the $t$-channel from the other production modes.

We remind the reader that the origin of this problem is solely due to the fact that we are performing a calculation with unstable particles that are kept stable, and the separation of the different processes is based on intermediate resonances, a procedure not well defined in quantum mechanics due to interference effects. On the contrary, signal regions in experimental analyses are defined via stable particles/objects (jets, $b$-jets, leptons, {\it etc.}) emerging from decays.
There are two kinds of possible strategies for addressing this problem. First, performing a calculation for the signature itself, as it has been done in Ref.~\cite{Frederix:2019ubd} for the NLO QCD+EW  corrections in $t$-channel single-top production or in Refs.~\cite{Frederix:2013gra, Cascioli:2013wga, Jezo:2016ujg}  for NLO QCD corrections in the $tW$ associated production, where in both cases  no  additional associated $Z$ or $H$ boson production has been taken into account. Second, adopt gauge-violating solutions based on the exclusion of specific diagrams, denoted in the literature as Diagram Removal (DR) \cite{Frixione:2008yi, Hollik:2012rc}, or gauge-invariant solutions based on the subtraction of on-shell contributions at the global or local level, denoted in the literature as Diagram Subtraction (DS) \cite{Beenakker:1996ch, Hollik:2012rc, Gavin:2013kga}.

The first kind of strategy implies an extremely challenging calculation for the case of single-top production with an extra emission of a Higgs boson, a $Z$ boson or an $\ell^+ \ell^-$ pair, especially for the case of EW corrections.\footnote{This would be equivalent to the calculation of Ref.~\cite{Frederix:2019ubd} with at least two additional particles in the final state. } The complexity of such a calculation is well beyond the state-of-the-art predictions available at the moment in the literature. Especially, this level of accuracy is not urgently needed. The second strategy could be in principle used in order to remove the contribution of $tW$ from the NLO corrections, namely, $tW$ production with the subsequent hadronic $W$ boson decay, see Fig.~\ref{fig:tW}. However, at variance with the case of the large $t \bar t$ contribution emerging in $tW$ at NLO QCD accuracy \cite{Frixione:2008yi, Demartin:2016axk}, the $tW$ cross section is in fact smaller than the single-top $t$-channel cross section. Moreover, signal regions in experimental analyses are not designed in order to suppress the $tW$ contribution.  For this reason, we believe that also the second strategy is not improving the theoretical predictions that are relevant for experimental analyses. Therefore, for the processes studied in this work, results obtained with no restrictions on the diagrams and the flavour of the jet in the final state should be preferred. From now on, the $tW$ associated production with  the subsequent hadronic $W$ boson  decay (see Fig.~\ref{fig:tW}) will be denoted as $tW_h$.

On the basis of the previous discussion, we now can motivate our strategy for evaluating flavour-scheme uncertainties. In particular, in the following we explain why we consider only differences among $t$-channel  contributions at NLO QCD accuracy for the flavour-scheme uncertainties of  $\NLOQCDEW$ predictions for which the $t$-channel selection is not performed.

The main problem is that, without separating the different productions modes, it is not straightforward to compare NLO QCD predictions for $tj$ production in the 4FS and in the 5FS. Indeed, in the 4FS, the $t$-channel process corresponds to the process $pp\TO t j \bar b$, which however includes also real emissions of gluons from $s$-channel diagrams. Unless a minimum $p_T$ for the light jet is required, these contributions are not infrared (IR) finite and therefore have to be excluded, both at LO and at NLO. On the other hand, the $s$-channel contribution not only is very small, but it also does not depend on the bottom PDF; its impact on the flavour-scheme choice is completely negligible. Thus, the $s$-channel contribution can be safely removed in the estimate of the flavour-scheme uncertainties. In principle, one may retain the contribution of $tW_h$ production arising from NLO corrections, however we believe it is preferable to exclude it, too. The reason is that the $tW_h$ contribution is of Born type both in the 4FS and in the 5FS. Therefore, being proportional to  $\alpha_s^2$ in the 4FS and to $\alpha_s$ in the 5FS, in the former case it entails a larger dependence on $\mu_R$. Moreover, in the 4FS also a larger dependence on $\mu_F$ is present, since large logarithms involving $m_b$ are not resummed. In other words, concerning the $tW_h$ contribution in our calculation, the 4FS simply leads to a less accurate prediction than the 5FS one. 
In conclusion, we use the $t$-channel only contributions for the estimate of flavour-scheme uncertainties. NLO EW corrections are not expected to play a major role in the flavour-scheme uncertainty and, as will also be shown in Sec.~\ref{sec:inclusive}, their impact on the scale uncertainties is below the percent level. Therefore, also for $\NLOQCDEW$ predictions, the estimate of flavour-scheme uncertainties is performed via the comparison of $\NLOQCD$ $t$-channel predictions.\footnote{Since very recently NLO EW corrections can be calculated also in the 4FS via {\sc MadGraph5\_aMC@NLO} \cite{Pagani:2020rsg}.} The quantities $(\NLOQCDEW^{5FS})^{+\Delta^{\rm 4-5FS}_+}_{-\Delta^{\rm 4-5FS}_-}$ and $(\NLOQCD^{5FS})^{+\Delta^{\rm 4-5FS}_+}_{-\Delta^{\rm 4-5FS}_-}$ that have been introduced in Sec.~\ref{sec:flavscheme} precisely correspond to the strategy that we have just outlined and, as already said,  will be denoted in Sec.~\ref{sec:results} as 5FS$_{\rm 4-5}^{\rm scale}$.

From the previous discussion it also becomes clear why the so-called ``multiplicative approach'' for the combination of NLO QCD and NLO EW corrections is not expected to lead to improved predictions. At variance with the standard additive procedure leading to the $\NLOQCDEW$ predictions, which is also denoted in the literature as ``additive approach'', in the multiplicative approach the $\NLO_2$ term is not only added on top of the $\NLOQCD$ predictions but it is also multiplied by the QCD $K$-factor, the $ (\LO_1+\NLO_1)/\LO_1$ ratio. The rationale behind this choice is the possibility of estimating mixed QCD--EW NNLO corrections of $\ord(\alpha_s \alpha)$ w.r.t.~the LO predictions and at the same time reduce the scale dependence of  the $\NLOQCDEW$ ones. However, the multiplicative approach is justified only when the dominant contributions from both the $\NLO_1$ and $\NLO_2$ terms factorise, the typical case being soft QCD corrections from the former and weak Sudakov logarithms from the latter. In our calculation both NLO QCD and NLO EW corrections induce the opening of a new production mechanisms, namely the $tW_h$ associated production. First, the $tW_h$ component in the NLO QCD corrections is not negligible, as also documented in Sec.~\ref{sec:results}. Second, separating the different production channels is not possible for the case of the EW corrections. For these reasons, in our study we have refrained from combining NLO QCD and NLO EW corrections in the multiplicative approach and we provide predictions only in the additive approach, $\NLOQCDEW$.

\subsection{Input Parameters}
\label{sec:input}

In this paper we provide results for proton--proton collisions at the LHC, with a centre-of-mass energy of 13 TeV.
In our calculation we use the following on-shell input parameters
\begin{align}
m_{Z}&=91.188~\textrm{GeV}\, ,&\quad  m_{W}&=80.385~\textrm{GeV}\, ,&\quad  m_H&=125~\textrm{GeV}\, ,& \nonumber \\
\Gamma_{Z}&=2.49707~\textrm{GeV}\, ,&\quad \Gamma_{W}&=2.09026~\textrm{GeV}\, ,& \Gamma_H&=0\, ,&\\
   m_{\textrm{t}}&=173.3~\textrm{GeV}\, ,&\quad m_{b}&=4.92~\textrm{GeV}\, ,&\quad \Gamma_{\textrm{t}}&=0\, , &\nonumber 
\end{align}
and employ the complex mass scheme \cite{Denner:1999gp,Denner:2005fg}. We have set $\gamh=0$ and $\Gamma_{t}=0$, since in our calculation there is always an external on-shell top quark and in the case of $tHj$ production an external on-shell Higgs boson. We also set $\Gamma_{Z}=0$ in the case of $tZj$ production. The value $m_{b}=4.92~\textrm{GeV}$ directly enters our calculations only in the 4FS. It has been chosen in order to be consistent with the value used in the PDF evolution of the PDF sets that we employ for the calculations in the 5FS. We discuss later in detail the PDF-set choice. We also note that while in the 5FS the Higgs boson does not couple to the $b$ quark, it does in the 4FS. However, the contribution of diagrams involving this coupling is subleading in the case of $tHj$ production, where  a top quark is present in the final state and a $W$-boson in the propagator, leading to much larger Higgs couplings. For this reason, in the 4FS we can safely use the on-shell $\mb$ value also for the bottom Yukawa interaction. See also Ref.~\cite{Demartin:2015uha} for more details. 
   
In our calculation EW interactions are renormalised in the $G_\mu$-scheme with 
\begin{equation}
    \gmu = 1.16639 \cdot 10^{-5} ~\gev^{-2}\,,
\end{equation}
while QCD interactions are renormalised  in the $\MSbar$-scheme with five active flavour in the 5FS and four active flavour in the 4FS. The numerical input and the $\mu_R$ dependence of $\alphas$ is directly taken from the PDF sets used in the calculation. In order to estimate QCD scale uncertainties, we vary independently by a factor of two $\mu_r$ and $\mu_f$ around the central value $\mu_0$ defined as follows,
\begin{align}
 \mu_0\equiv H_T/6&=\frac{\sum_i m_{T,i}}{6}\,,~~~  i=t,H,b&{\rm for~}tHj \, ,   \label{eq:scaleH}    \\ 
 \mu_0\equiv H_T/6&=\frac{\sum_i m_{T,i}}{6}\,,~~~ i=t,Z,b&{\rm for~}tZj  \, , \label{eq:scaleZ}    \\
 \mu_0\equiv H_T/6&=\frac{\sum_i m_{T,i}}{6}\,,~~~  i=t,Z(\ell^+\ell^-),b&{\rm for~} \tllj \, .  \label{eq:scalell}    
\end{align}
 The scale definition in Eq.~\eqref{eq:scaleH} is based on the findings of Refs.~\cite{Maltoni:2012pa, Demartin:2015uha} and the same rationale has been used for Eqs.~\eqref{eq:scaleZ} and \eqref{eq:scalell}. With $t$ and $b$ we refer to both quarks and antiquarks, and the momentum of the bottom (anti)quark is set to zero when this particle is not present in the final state\footnote{Since in our calculation there are not $\gamma,g\TO b \bar b$ splittings in the final state, this definition is IR safe.}. More details about the scale dependence will be discussed in Sec.~\ref{sec:4and5inclusive}.  \\

 The choice of the PDF set is motivated by a few aspects that are explained in the following. 
First, our calculation is performed at NLO QCD+EW accuracy and therefore (at least) the same level of accuracy has to be present for the PDFs themselves. Second, in order to evaluate flavour-scheme uncertainties,  both a 4FS and a 5FS version of the same PDF fit have to be available.
Therefore, the only three possible options at the moment are: {\sc\small NNPDF3.0}  \cite{Ball:2014uwa, Bertone:2016ume}, {\sc\small NNPDF3.1} \cite{Ball:2017nwa, Bertone:2017bme} and {\sc\small MMHT2014/MMHT2015} \cite{Harland-Lang:2014zoa, Harland-Lang:2019pla}. All these three sets  are  accurate up to NNLO in QCD and NLO in QED accuracy  and include a photon density based on the   {\sc\small LUXqed} parameterisation \cite{Manohar:2016nzj, Manohar:2017eqh}. 
The set {\sc\small NNPDF3.1} should be preferred over the {\sc\small NNPDF3.0} one, being an improvement of the former, and we choose it for our calculations. Notably, in the case of $b$-initiated processes this improvement cannot be neglected. We have verified that results in the 5FS obtained with {\sc\small NNPDF3.1} and  {\sc\small NNPDF3.0} at NLO QCD accuracy are not compatible within their PDF uncertainties;  the difference between   them is several times larger than the respective PDF uncertainties. These differences have to be attributed to the different numerical input values for $m_b$ in the {\sc\small NNPDF3.0} and {\sc\small NNPDF3.1} PDF fits,\footnote{We explicitly verified that these effects originate from the different value of the mass of the bottom quark via the  {\sc\small NNPDF2.1} PDF sets \cite{Ball:2011mu}, which allows to use different values for $m_b$; consistent deviations have been found. } which can induce large effects on the bottom PDF and in turn to the bottom--gluon luminosity, entering the LO predictions for all the processes considered in this paper. Especially, in our calculation we set $\mu_0=H_T/6$, which is quite small, and the smaller  the factorisation scale, the larger  the effect induced by a different $\mb$ value, since  $\mb$ determines the threshold condition for the bottom-quark PDF.  On the other hand, this effect is smaller if instead of NLO PDFs one employs PDFs at NNLO accuracy. For this reason we suggest to avoid to use of NLO PDFs for the calculation of the processes considered in this work and we adopt NNLO PDFs. We also note that with this choice {\sc\small NNPDF3.1}  and {\sc\small MMHT2014/MMHT2015} predictions  are very well compatible. As a last remark we want also point out that, to the best of our knowledge, no 4FS PDF set  including a photon PDF and NLO QED effects is available at the moment, but would be necessary for NLO EW corrections in the 4FS. 

Finally, we describe the clustering procedure that we perform in order to obtain jets and dressed leptons. First of all we recombine possible photons that are present in the final state, due to NLO EW corrections or shower effects, with leptons. In fact, this step concerns only the $\tllj$ process. A dressed lepton is  obtained by recombining a bare lepton $\ell$ with any photon $\gamma$ satisfying the condition 
\begin{equation}
\Delta R(\ell, \gamma) < 0.1\,, \label{eq:recQED}
\end{equation}
 where $\Delta R(\ell, \gamma) \equiv \sqrt{(\Delta \eta(\ell, \gamma))^2+(\Delta \phi(\ell, \gamma))^2} $ and  $\eta(\ell, \gamma)$ and $\Delta \phi(\ell, \gamma)$ are the difference of the bare-lepton and photon pseudorapidities and azimuthal angles, respectively. In case that the condition  \eqref{eq:recQED} is satisfied for both $\ell^+$ and $\ell^-$, the photon is clustered together with the bare lepton for which $\Delta R(\ell, \gamma)$ is the smallest.  After this, we cluster jets via the anti-$k_T$ algorithm \cite{Cacciari:2008gp}  as implemented in {\sc \small FastJet}~\cite{Cacciari:2011ma} using the parameters 
\begin{equation}
p_T^{\rm min}=25~{\rm GeV}\,, \qquad R=0.5\,, 
\end{equation}
and including also the previously unrecombined photons in the clustering procedure. This means that in our calculation, especially at fixed order, a jet can correspond to a single photon.\footnote{In many LHC analyses jets are defined with up to 99\% of their energy of electromagnetic origin and even up to 90\% that can be associated to a single photon. More details can be found in  Ref.~\cite{Frederix:2016ost}.} However, it is important to note that in this work the jet definition is relevant only for differential distributions and not for total cross sections. Indeed, the $tHj$, $tZj$ and $\tllj$ processes are all properly defined and IR finite without tagging any jet. When we will consider $b$-jets, we will simply refer to jets containing a bottom (anti)quark, without any restriction on their pseudorapidity. Also, since in our calculation there are no $\gamma,g\TO b \bar b$ splittings in the final state, $b$-jets cannot include more than one bottom (anti)quark and no IR safety problems are present in their definition also in the 5FS.

  \section{Numerical Results}
 \label{sec:results}

In this section we present and discuss the numerical results of our study. We start with results concerning  total cross sections, Sec.~\ref{sec:inclusive}, and afterwards we comment in detail on the case of differential distributions in Sec.~\ref{sec:differential}. In both cases, following the strategy described in Sec.~\ref{sec:flavscheme}, we compare  4FS and 5FS results in order to evaluate flavour-scheme and scale uncertainties and then we quantify and discuss the impact of electroweak corrections.
For the $\tllj$ process we show results for two different selection cuts on the invariant mass of the lepton pair $m(\ell^+ \ell^-)$:
\begin{enumerate}
\item $m(\ell^+ \ell^-)>30$ GeV, dubbed as  ``inclusive'',
\item  $|m(\ell^+ \ell^-)-m_Z| <10$  GeV, dubbed as $Z$-peak. 
\end{enumerate}
The first choice is inspired by the experimental 
measurements of Refs.~\cite{Sirunyan:2018zgs,Aad:2020wog}, which report unfolded results for this kinematic region. The second choice is 
motivated by the experimental analysis of Ref.~\cite{Aad:2020wog}, which applies this requirement when selecting the events. 
Finally, in Sec.~\ref{sec:shower}, we discuss the impact of the parton shower, including or not QED effects, on top of NLO QCD predictions for the specific case of $\tllj$ production.

All results in this section have been obtained via the {\sc\small MadGraph5\_aMC@NLO} framework \cite{Alwall:2014hca}. Results including NLO QCD and EW corrections employ the latest version of {\sc\small MadGraph5\_\-aMC@NLO} \cite{Frederix:2018nkq}, which is publicly available and allows to calculate NLO EW corrections, and more in general predictions at complete-NLO accuracy, for any SM process. The  {\sc\small MadGraph5\_aMC@NLO} framework \cite{Alwall:2014hca} deals with IR singularities via the so-called FKS method~\cite{Frixione:1995ms,
Frixione:1997np}, which has been automated in \mf~\cite{Frederix:2009yq,
Frederix:2016rdc}. One-loop amplitudes are evaluated via 
different types of integral-reduction techniques, namely,  the  OPP method~\cite{Ossola:2006us} or
 the Laurent-series expansion~\cite{Mastrolia:2012bu},
and  techniques for tensor-integral reduction~\cite{Passarino:1978jh,Davydychev:1991va,Denner:2005nn}.
The module \ml~\cite{Hirschi:2011pa}, which is employed for generating  the amplitudes, automates these techniques and switches dynamically among them. We remind the reader that
 the codes \ct~\cite{Ossola:2007ax}, \nin~\cite{Peraro:2014cba,
Hirschi:2016mdz} and \coll~\cite{Denner:2016kdg} are employed within \ml, which also includes an in-house
implementation of the {\sc OpenLoops} optimisation~\cite{Cascioli:2011va}.

\subsection{Inclusive results}
\label{sec:inclusive}
\subsubsection{QCD scale uncertainties in the 4FS and 5FS}
\label{sec:4and5inclusive}

For the determination of scale and flavour-scheme uncertainties we follow the strategy that has been described in Sec.~\ref{sec:flavscheme}. Therefore, according to this strategy, in this section we focus on 4FS and 5FS predictions for the $t$-channel contributions to the $tHj$, $tZj$ and $\tllj$ processes. 
In analogy with Ref.~\cite{Demartin:2015uha}, which focuses on $tHj$ production, we  analyse the 4FS and 5FS scale dependence at LO and NLO in QCD for the four processes that we consider. We use the setup described in the previous section and to this purpose we vary the central value $\mu_0$ for the  renormalisation 
 and factorisation scale, which has been defined in Eq.~\eqref{eq:scalell}, up and down by a factor of 8.
 In particular, in the main panel of each of the plots  in Fig.~\ref{fig:fig3rep}, the solid lines correspond  to the case $\mu_R=\mu_F=r_{\mu} \mu_0$, where $1/8<r_{\mu}<8$. At NLO,  we also explore the impact of  non-equal $\mu_R$ and $\mu_F$ values (off-diagonal variation).
 For a given $\mu_R=\mu_F$, the coloured bands in Fig.~\ref{fig:fig3rep} show the range of cross sections obtained by either keeping $\mu_R$ or $\mu_F$ fixed at $r_{\mu} \mu_0$ and moving the other one. In both cases, the variation is performed in the range $1/2<\mu_F/\mu_R<2$.\footnote{This is equivalent to the 7-point variation around the central value $\mu_R=\mu_F=r_{\mu} \mu_0$. }
In each of the plots  of Fig.~\ref{fig:fig3rep} we also show the QCD $K$-factor, namely the ratio between the $\NLOQCD$ and LO predictions, in the lower inset, for both the 4FS and the 5FS.

\begin{figure}[t]
\centering
\includegraphics[width=0.425\textwidth]{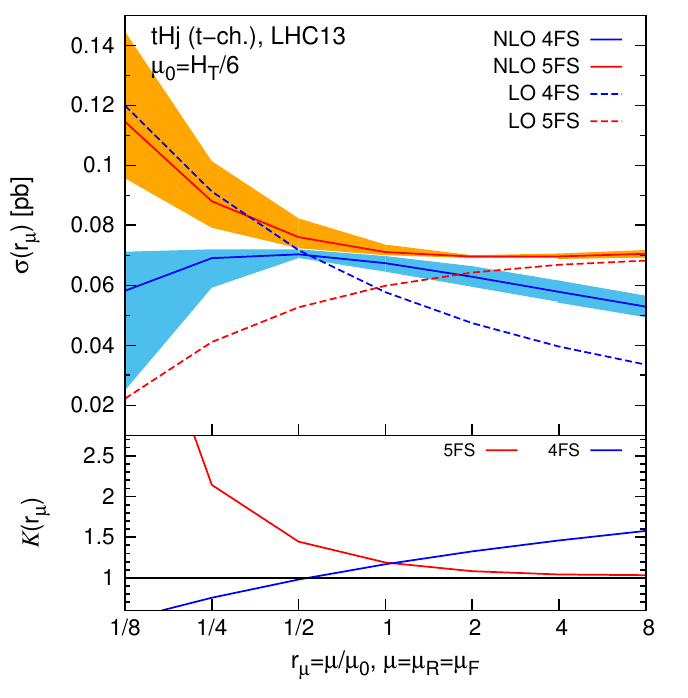}
\includegraphics[width=0.425\textwidth]{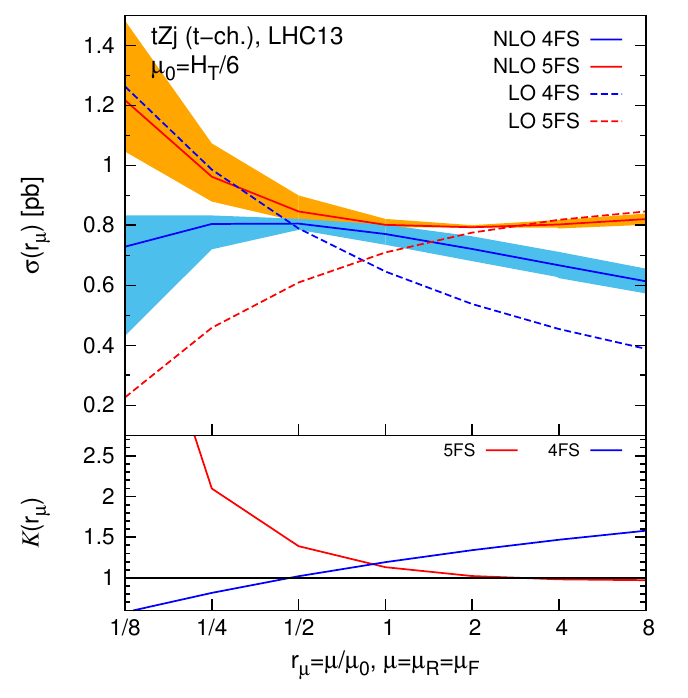} \\
\includegraphics[width=0.425\textwidth]{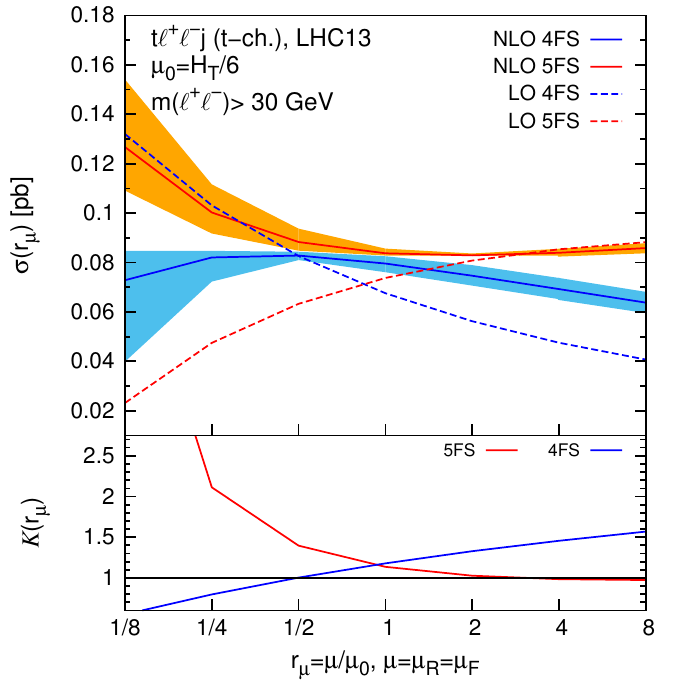}
\includegraphics[width=0.425\textwidth]{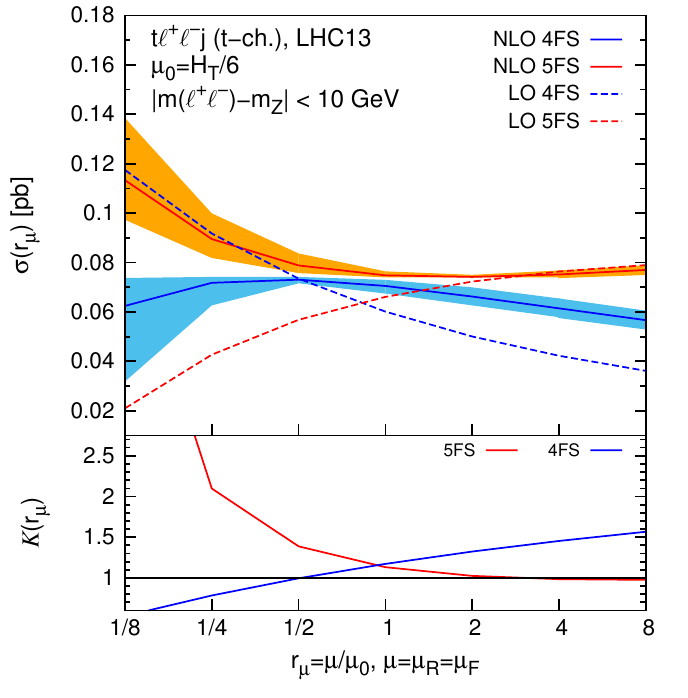}
\caption{Scale dependence of total cross sections for $tHj$ production (top-left), $tZj$ production (top-right) and $\tllj$ production for the ``inclusive'' (bottom-left) and $Z$-peak (bottom-right) case in the 4FS and 5FS.} 
\label{fig:fig3rep}       
\end{figure}

The first observation is that all four processes behave in a very similar way and the following discussion applies to all of them. NLO QCD corrections reduce
 the scale uncertainties for both the 4FS and 5FS predictions. The difference between the two schemes is minimised in the region of 
 $\mu_0/2<\mu<\mu_0$, with the two uncertainty bands touching each other.  In  this scale region, the difference between the central values lies in the 5\% ballpark.  The 5FS $K$-factor strongly increases at low scales and approaches the value of 1 at high  scales, whilst in the 4FS QCD corrections decrease 
 the cross section at low scales ($K$-factor$\,<1$).  Given the flatness of the NLO plots shown in Fig.~\ref{fig:fig3rep} and how narrow the
  bands of the off-diagonal variation are, it is clear that if we consider only the 4FS or 5FS with the typical scale variation
  of a factor of two up and down, we will obtain very small uncertainties. These scale uncertainties will not be large enough to enclose both 
  the 4FS and 5FS central values. Therefore, the combination of 4FS and 5FS uncertainties, as described in the previous section, is necessary in order to properly account for missing higher-order QCD effects.

\subsubsection{NLO QCD+EW predictions}
We proceed to the computation of  total cross sections at NLO QCD+EW accuracy, without selecting $t$-channel diagrams; $s$-channel and $tW_h$ contributions are retained as explained in Sec.~\ref{sec:ch-sepa}.
Inclusive results for the processes that we consider in this work, $tHj$, $tZj$ and $\tllj$, are shown in Tab.~\ref{tab:comp_xsec}, using the settings described in Sec.~\ref{sec:setup}.   The two dilepton invariant mass cuts for $\tllj$  will allow us to investigate 
the impact of EW corrections and compare this to the result for the undecayed $tZj$ process. 

\begin{table}[h]
\renewcommand{\arraystretch}{2.5}
\scriptsize
\begin{center}
\begin{tabular}{c c c | c c}

Accuracy & Channel & FS & $t H j$ & $t Z j$  \\
\hline
\multirow{3}{*}{$\NLOQCD$} & \multirow{3}{*}{$t$-ch.} & 4FS & $68.1(1)_{-4.5 (-6.6  \%)}^{+2.7 (+4.0  \%)}~_{-0.4 (-0.5  \%)}^{+0.4 (+0.5  \%)}$  & $764(1)_{-48 (-6.2  \%)}^{+33 (+4.3  \%)}~_{-3 (-0.4  \%)}^{+3 (+0.4  \%)}$ \\
 &  & 5FS & $71.3(1)_{-1.7 (-2.4  \%)}^{+5.2 (+7.2  \%)}~_{-0.3 (-0.5  \%)}^{+0.3 (+0.5  \%)}$ & $805(1)_{-8 (-1.0  \%)}^{+45 (+5.5  \%)}~_{-3 (-0.4  \%)}^{+3 (+0.4  \%)}$ \\
 &  & 5FS$_{\rm 4-5}^{\rm scale}$ & $71.3(1)_{-7.7 (-10.9  \%)}^{+5.2 (+7.2  \%)}~_{-0.3 (-0.5  \%)}^{+0.3 (+0.5  \%)}$ & $805(1)_{-89 (-11.1  \%)}^{+45 (+5.5  \%)}~_{-3 (-0.4  \%)}^{+3 (+0.4  \%)}$ \\
\hline
\multirow{2}{*}{$\NLOQCD$} & \multirow{2}{2cm}{ \centering $t$-ch., $s$-ch., $tW_h$} & 5FS & $85.1(2)_{-2.3 (-2.7  \%)}^{+5.4 (+6.4  \%)}~_{-0.5 (-0.6  \%)}^{+0.5 (+0.6  \%)}$ & $895(2)_{-16 (-1.8  \%)}^{+46 (+5.1  \%)}~_{-4 (-0.4  \%)}^{+4 (+0.4  \%)}$ \\
 &  & 5FS$_{\rm 4-5}^{\rm scale}$ & $85.1(2)_{-9.2 (-10.9  \%)}^{+6.2 (+7.2  \%)}~_{-0.5 (-0.6  \%)}^{+0.5 (+0.6  \%)}$ & $895(2)_{-99 (-11.1  \%)}^{+50 (+5.5  \%)}~_{-4 (-0.4  \%)}^{+4 (+0.4  \%)}$ \\
\multirow{2}{*}{$\NLOQCDEW$} & \multirow{2}{2cm}{ \centering $t$-ch., $s$-ch., $tW_h$} & 5FS & $82.2(2)_{-2.4 (-2.9  \%)}^{+5.6 (+6.8  \%)}~_{-0.5 (-0.6  \%)}^{+0.5 (+0.6  \%)}$ & $904(2)_{-19 (-2.1  \%)}^{+42 (+4.7  \%)}~_{-4 (-0.4  \%)}^{+4 (+0.4  \%)}$ \\
 &  & 5FS$_{\rm 4-5}^{\rm scale}$ & $82.2(2)_{-8.9 (-10.9  \%)}^{+5.9 (+7.2  \%)}~_{-0.5 (-0.6  \%)}^{+0.5 (+0.6  \%)}$ & $904(2)_{-100 (-11.1  \%)}^{+50 (+5.5  \%)}~_{-4 (-0.4  \%)}^{+4 (+0.4  \%)}$ \\
\\
Accuracy & Channel & FS & $t \ell^+ \ell^- j$  (``inclusive'') & $t \ell^+ \ell^- j$  ($Z$-peak)  \\

\hline
\multirow{3}{*}{$\NLOQCD$} & \multirow{3}{*}{$t$-ch.} & 4FS & $80.2(2)_{-5.0 (-6.2  \%)}^{+3.7 (+4.6  \%)}~_{-0.3 (-0.4  \%)}^{+0.3 (+0.4  \%)}$ & $70.9(2)_{-4.4 (-6.2  \%)}^{+3.1 (+4.3  \%)}~_{-0.3 (-0.4  \%)}^{+0.3 (+0.4  \%)}$ \\
 &  & 5FS & $84.0(1)_{-0.9 (-1.0  \%)}^{+4.7 (+5.6  \%)}~_{-0.3 (-0.4  \%)}^{+0.3 (+0.4  \%)}$ & $75.0(1)_{-0.8 (-1.0  \%)}^{+4.2 (+5.6  \%)}~_{-0.3 (-0.4  \%)}^{+0.3 (+0.4  \%)}$ \\
 &  & 5FS$_{\rm 4-5}^{\rm scale}$ & $84.0(1)_{-8.7 (-10.4  \%)}^{+4.7 (+5.6  \%)}~_{-0.3 (-0.4  \%)}^{+0.3 (+0.4  \%)}$ & $75.0(1)_{-8.5 (-11.3  \%)}^{+4.2 (+5.6  \%)}~_{-0.3 (-0.4  \%)}^{+0.3 (+0.4  \%)}$ \\
\hline
\multirow{2}{*}{$\NLOQCD$} & \multirow{2}{2cm}{ \centering $t$-ch., $s$-ch., $tW_h$} & 5FS & $93.7(2)_{-1.7 (-1.8  \%)}^{+4.9 (+5.2  \%)}~_{-0.4 (-0.4  \%)}^{+0.4 (+0.4  \%)}$ & $83.4(2)_{-1.5 (-1.8  \%)}^{+4.3 (+5.1  \%)}~_{-0.4 (-0.4  \%)}^{+0.4 (+0.4  \%)}$ \\
 &  & 5FS$_{\rm 4-5}^{\rm scale}$ & $93.7(2)_{-9.7 (-10.4  \%)}^{+5.2 (+5.6  \%)}~_{-0.4 (-0.4  \%)}^{+0.4 (+0.4  \%)}$ & $83.4(2)_{-9.4 (-11.3  \%)}^{+4.6 (+5.6  \%)}~_{-0.4 (-0.4  \%)}^{+0.4 (+0.4  \%)}$ \\
 \multirow{2}{*}{$\NLOQCDEW$} & \multirow{2}{2cm}{ \centering $t$-ch., $s$-ch., $tW_h$} & 5FS & $89.6(2)_{-1.7 (-1.9  \%)}^{+5.1 (+5.7  \%)}~_{-0.4 (-0.4  \%)}^{+0.4 (+0.4  \%)}$ & $77.2(2)_{-1.5 (-1.9  \%)}^{+4.9 (+6.3  \%)}~_{-0.3 (-0.4  \%)}^{+0.3 (+0.4  \%)}$ \\
 &  & 5FS$_{\rm 4-5}^{\rm scale}$ & $89.6 (2)_{-9.3 (-10.4  \%)}^{+5.0 (+5.6  \%)}~_{-0.4 (-0.4  \%)}^{+0.4 (+0.4  \%)}$ & $77.2(2 )_{-8.7 (-11.3  \%)}^{+4.3 (+5.6  \%)}~_{-0.3 (-0.4  \%)}^{+0.3 (+0.4  \%)}$ \\
\end{tabular}
\end{center}
\caption{Total cross-section for $tHj$, $tZj$ and $\tllj$ production. The uncertainties are scale and PDF of the form $\pm$ absolute ($\pm$ relative in \%). The first number in parentheses after the central value is the absolute statistical error.}
\label{tab:comp_xsec}  
\end{table}

For each process, in the first block we show results for the $t$-channel mode in the 4FS and 5FS at NLO in QCD. The 4FS and 5FS combined results, denoted as 5FS$_{\rm 4-5}^{\rm scale}$, are obtained from the combination of the 4FS and 5FS uncertainties as described in detail in Sec.~\ref{sec:flavscheme}. In the second block we show the $\NLOQCD$ and $\NLOQCDEW$ results in the 5FS including all the contributions ($t$-ch., $s$-ch., and $tW_h$-assoc.). In both cases we show first the pure 5FS result and then the 5FS$_{\rm 4-5}^{\rm scale}$ result. The latter is obtained using the 5FS central value, but now assigning as scale uncertainty the rescaled scale-uncertainty from the NLO QCD combination between 4FS and 5FS in the $t$-channel only case, the result in the third line of the first block. The $\NLOQCDEW$ prediction in the 5FS$_{\rm 4-5}^{\rm scale}$ is  at the moment the most precise and accurate prediction and should be taken as reference value for $tHj$, $tZj$ and $\tllj$ production. For a detailed discussion of the motivations and the procedure for assigning the scale and flavour-scheme uncertainties, see Secs.~\ref{sec:flavscheme} and \ref{sec:ch-sepa}. Concerning the PDF uncertainties, they are also reported in Tab.~\ref{tab:comp_xsec} and always refer to the central value.

QCD and EW $K$-factors are reported in Tab.~\ref{tab:K_factors}, both for the $t$-channel only case and including all the contributions. Specifically, we show the $\NLOQCD/\LO$ and the $\NLOQCDEW/\NLOQCD$ ratios, the former both in the 4FS and 5FS, the latter only in the 5FS.
\begin{table}[h]
\renewcommand{\arraystretch}{1.8}
\normalsize
\begin{center}
\begin{tabular}{c c c | c c}
FS & Channel & $K$-factor & $t H j$ & $t Z j$  \\
\hline
 4FS & \multirow{2}{2cm}{ \centering $t$-ch.} & \multirow{2}{*}{$\NLOQCD/\LO$} & 1.17 & 1.18 \\
 5FS & & & 1.20 & 1.13 \\
\hline
 \multirow{2}{*}{5FS} & \multirow{2}{2cm}{ \centering $t$-ch., $s$-ch., $tW_h$} & $\NLOQCD/\LO$ & 1.37 & 1.24 \\
  & & $\NLOQCDEW/\NLOQCD$ & 0.97 & 1.01 \\
\\
FS & Channel & $K$-factor & $t \ell^+ \ell^- j$  (``inclusive'') & $t \ell^+ \ell^- j$  ($Z$-peak)  \\
\hline
 4FS & \multirow{2}{2cm}{ \centering $t$-ch.} & \multirow{2}{*}{$\NLOQCD/\LO$} & 1.18 & 1.18 \\
 5FS & & & 1.13 & 1.13 \\
\hline
 \multirow{2}{*}{5FS} & \multirow{2}{2cm}{ \centering $t$-ch., $s$-ch., $tW_h$} & $\NLOQCD/\LO$ & 1.24 & 1.24 \\
  & & $\NLOQCDEW/\NLOQCD$ & 0.96 & 0.93 \\
\end{tabular}
\end{center}
\caption{QCD and EW $K$-factors for all processes. The statistical error is beyond the digits displayed here.}
\label{tab:K_factors}  
\end{table}
Several observations are in order. As already discussed, scale uncertainties of the $\NLOQCD$ results are quite small, reaching at most 7\%, for the individual 4FS and 5FS predictions for all four processes considered here. At the same time PDF uncertainties remain below the percent level. On the other hand, the $t$-channel results differ by about 4\% between the 4FS and 5FS.  Combining the 4FS and 5FS scale variations enlarges the scale uncertainty to at most 11\% in the lower direction, in order to encompass the lower edge of the 4FS uncertainty band.
Including the $s$-channel and $W$-associated channel increases the cross section by 12\% for $tZj$ and $\tllj$ and 19\% for $tHj$. We notice that NLO QCD scale uncertainties for the pure 5FS results are at the same level with and without the selection of the $t$-channel modes. This fact supports our strategy for the evaluation of flavour-scheme and scale uncertainties.

Electroweak corrections have a different impact on the four processes considered. They decrease the $\NLOQCD$ $tHj$ cross section by 3\%, increase the $tZj$ one by 1\% whilst the $Z$-peak results and ``inclusive'' results are reduced by 7\% and 4\% respectively. The presence of the $Z\TO \ell^+ \ell^-$ decay has a non-negligible impact on the relative size  of EW corrections. Indeed, the radiation of photons from the leptons induces the migration of events outside the region $m(\ell^+ \ell^-)\sim m_Z$. This is the reason why in the $Z$-peak case  NLO EW corrections are larger in magnitude than in the ``inclusive'' case: more events migrate outside the selected phase-space region.  Nevertheless, for all the processes and cuts considered, the size of EW corrections is smaller than the combined 5FS$_{\rm 4-5}^{\rm scale}$  scale uncertainty band. Scale uncertainties of the $\NLOQCDEW$ predictions are as expected similar to the $\NLOQCD$ ones. We want to point out that, with the exception of the $tZj$ case, if we did not combine 4FS and 5FS scale uncertainties, the $\NLOQCDEW$ central values would have been outside the $\NLOQCD$ scale-uncertainty band.

Comparing the $tZj$ and $\tllj$ results, we notice that both the $Z$-peak and ``inclusive'' results differ from what one would naively expect from the narrow-width approximation $\sigma(\tllj)=\sigma(tZj)\times Br(Z\to \ell^+\ell^-)$ as there is a significant contribution from off-shell effects and the photon contribution. It is also worth mentioning that moving away from the $Z$-peak  and allowing a looser selection on the lepton pair invariant mass, as done for the ``inclusive'' case,  the cross section increases by more than 15\%. 

\subsection{Differential results}
\label{sec:differential}
\subsubsection{QCD scale uncertainties in the 4FS and 5FS}

\begin{figure}[!t]
\centering
\includegraphics[width=0.315\textwidth]{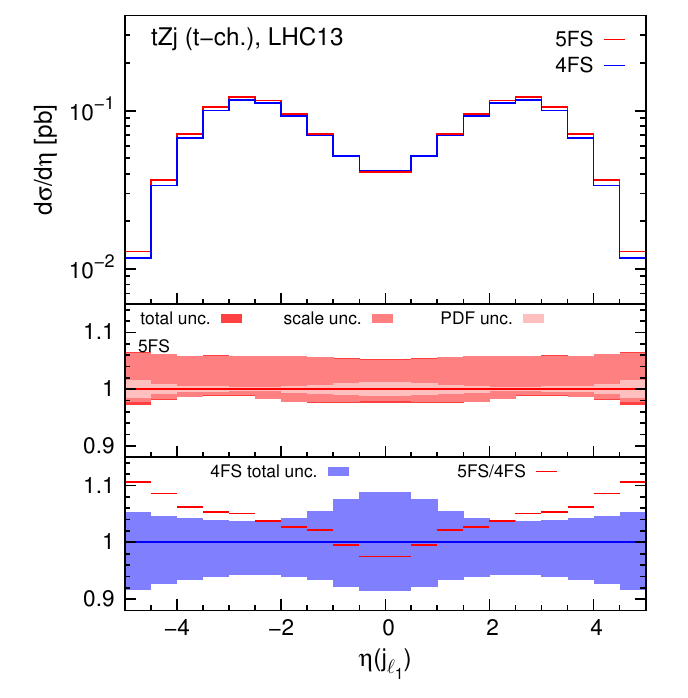}
\includegraphics[width=0.315\textwidth]{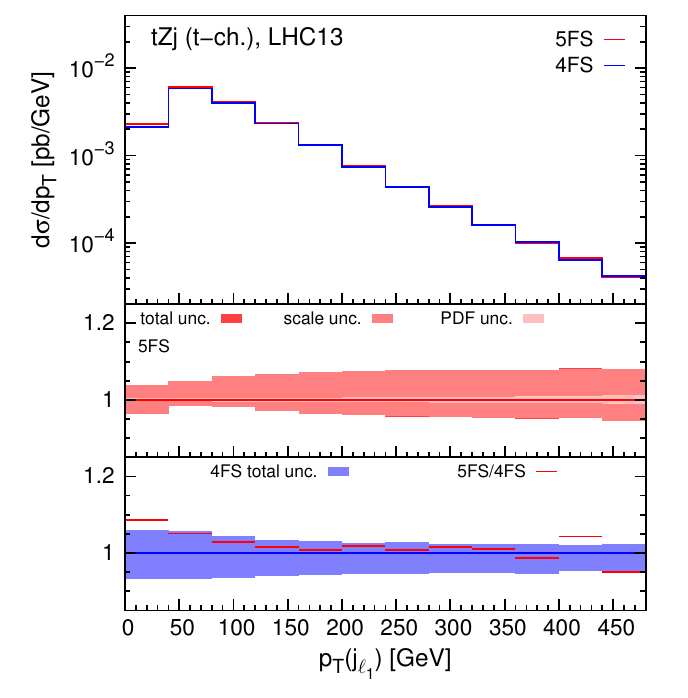}\\
\includegraphics[width=0.315\textwidth]{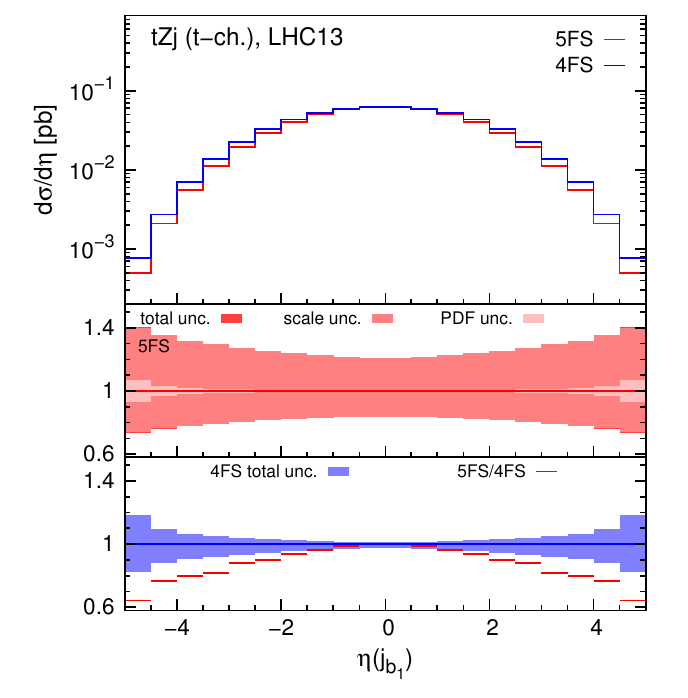}
\includegraphics[width=0.315\textwidth]{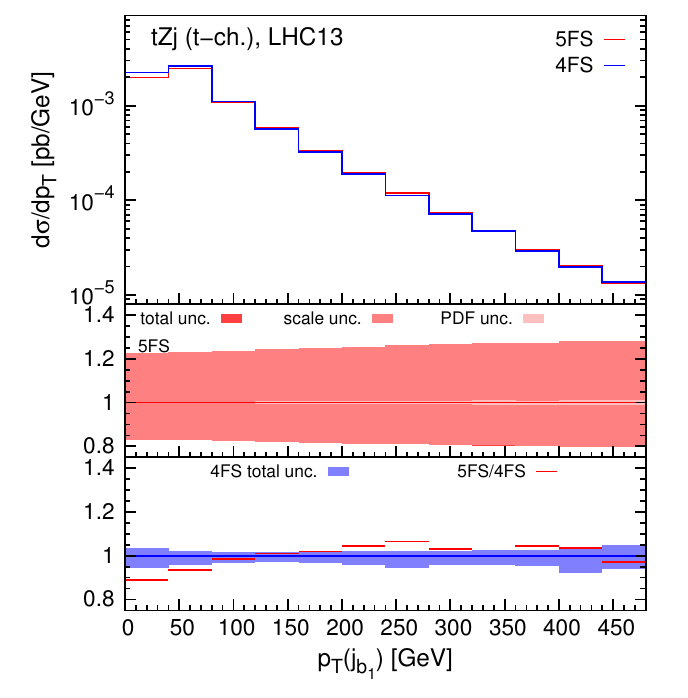}\\
\includegraphics[width=0.315\textwidth]{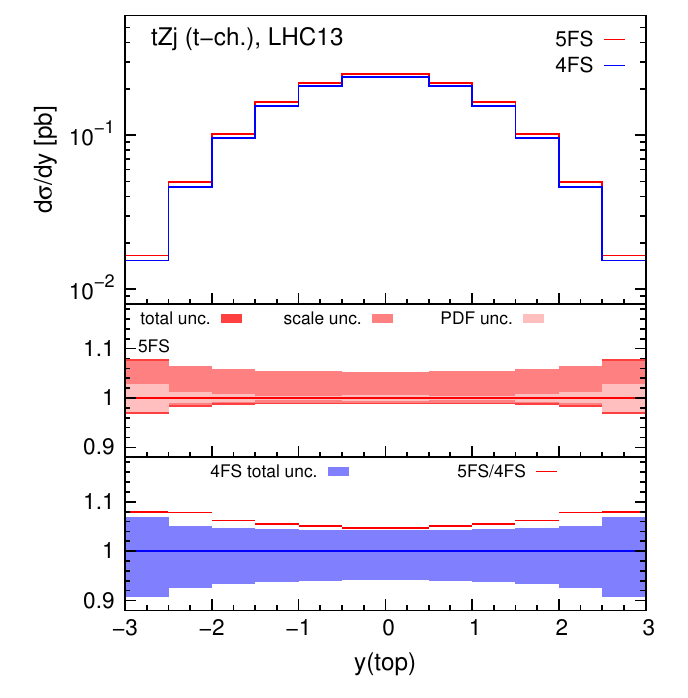}
\includegraphics[width=0.315\textwidth]{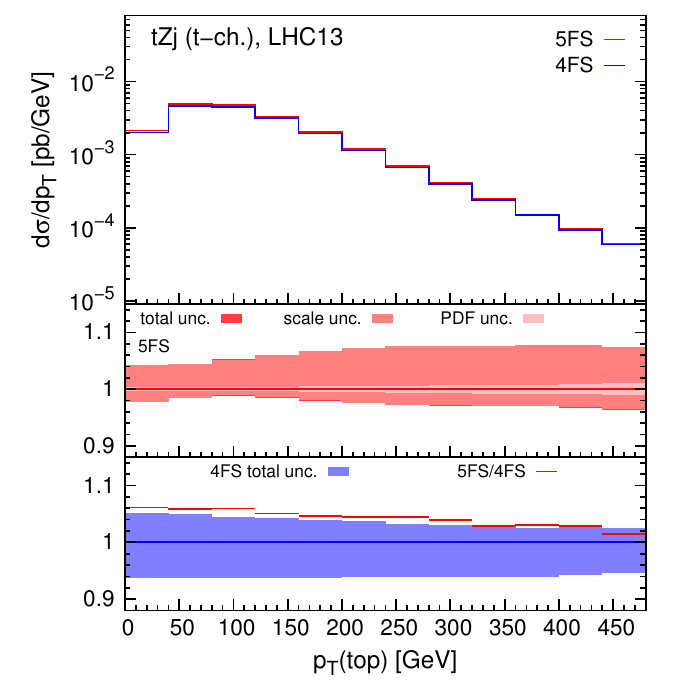}\\
\includegraphics[width=0.315\textwidth]{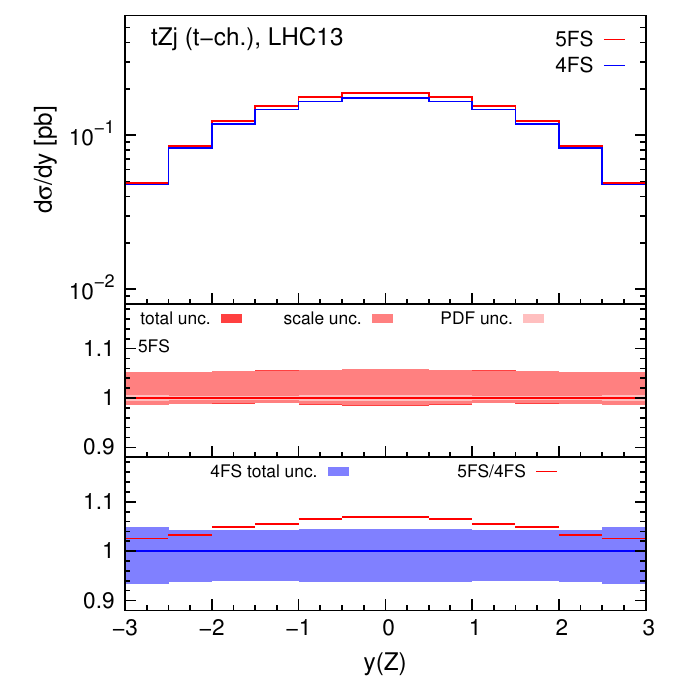}
\includegraphics[width=0.315\textwidth]{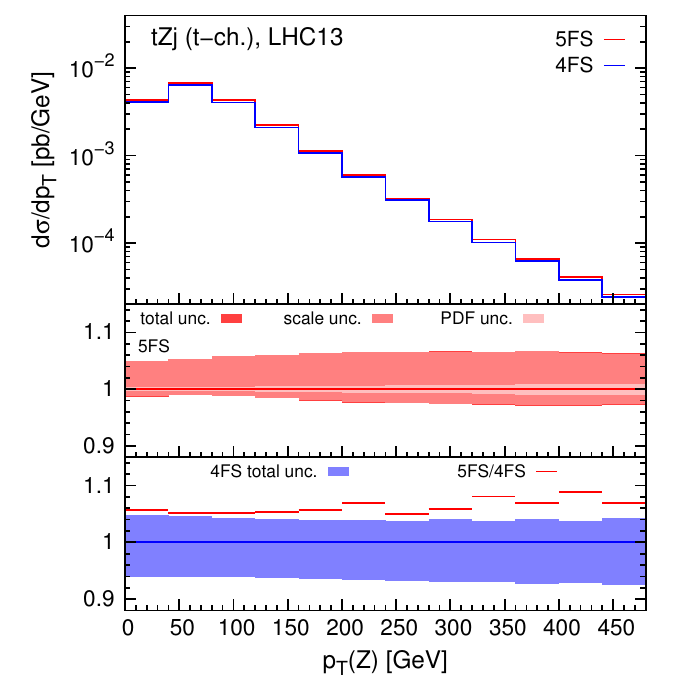}
\caption{ Comparison between 4FS and 5FS for $t$-channel $tZj$ at NLO QCD. Rapidity and transverse momentum distributions are shown for the hardest light jet, the hardest $b$-tagged jet, the top quark and the $Z$ boson. } 
\label{fig:tZj_4vs5FS}
\end{figure}

In order to quantify the differences between the 4FS and 5FS at a differential level we consider several key observables. 
As a detailed comparison of 4FS and 5FS predictions for $tHj$ production has already been performed in Ref.~\cite{Demartin:2015uha}, here
our main focus is on $tZj$ and $\tllj$ production, but one should note that the same qualitative behaviour is observed for both $tHj$ and $tZj$ production.
Moreover, since the impact of NLO QCD corrections is almost identical for $tZj$ and $\tllj$ production, in the context of 4FS and 5FS comparisons at the differential level we explicitly consider only $tZj$ production. Again, following the strategy already employed for total rates and explained in Secs.~\ref{sec:flavscheme} and \ref{sec:ch-sepa}, in this context we consider $t$-channel only predictions. We recall that the scale-uncertainty bands for the 5FS$_{\rm 4-5}^{\rm scale}$ predictions, which will be discussed in the next section, correspond to the bin-by-bin envelope of precisely the 4FS and 5FS scale uncertainty bands that we are going to  show in this section.

In Fig.~\ref{fig:tZj_4vs5FS} we show $\NLOQCD$ results for the transverse momentum ($p_T$) and rapidity or pseudorapidity ($y$ or $\eta$) of the hardest light jet ($j_{l,1}$), the 
hardest $b$-jet ($j_{b,1}$), top quark and $Z$ boson. In each plot, we show 5FS and 4FS predictions in the main panel. In the first inset we show scale and PDF uncertainties in the 5FS, summed in quadrature and normalised to the corresponding central value. Instead, in the second inset, we show scale uncertainties in the 4FS,  again normalised to the corresponding central value, together with the ratio of the 5FS and 4FS predictions.

 The largest difference between the 4FS and 
5FS is observed for the $b$-jet pseudo-rapidity distribution, reaching up to 35\% in the high rapidity region, which is however beyond the reach of realistic experimental range for $b$-jet tagging in ATLAS and CMS. This effect is due to the fact that the $b$-jet is more 
central in the 5FS computation, as expected and also observed in single top production \cite{Campbell:2009ss}. It is actually remarkable, how the $\NLOQCD$ predictions in the 4FS, which involves a $b$-jet already at LO, and in the 5FS, which involves a $b$-jet only at $\NLOQCD$ for this process, are close in value.\footnote{We have investigated this aspect and found that NLO QCD corrections are small in the 4FS. Thus, this comparison can also be viewed as NLO QCD in the 5FS versus LO in the 4FS, for which a similar $p_T(j_{b,1})$ spectrum is expected, especially when  $|\eta(j_{b,1})|$ is not large. } 
The hardest light jet is more peripheral and the $Z$ boson more central in the 5FS, but the differences with the corresponding 4FS predictions  never exceed 10\%. Looking at the transverse momentum distributions we 
find no striking differences in the shapes for the 4FS and 5FS. Scale uncertainties are similar in size for the top, light jet and $Z$-boson observables when comparing 4FS and 
5FS. For the $b$-jet however the scale uncertainties are significantly smaller for the 4FS as a $b$-quark is present 
already at LO and therefore $b$-jet observables are computed at NLO QCD accuracy. In the 5FS $b$-quarks emerge only at 
NLO and therefore $b$-jet observables are effectively described at LO accuracy.

We note here that whilst the qualitative behaviour of $tHj$ and $tZj$ is similar, the differences listed above 
between 4FS and 5FS are more pronounced for $tHj$. In particular, the $b$-jet transverse-momentum distribution is 
significantly harder for the 5FS. The presence of differences between the same differential distribution for these two processes is not surprising, since although $tHj$ and $tZj$ are similar processes, they receive different contributions. For instance, the $Z$ boson couples to all particles involved in $tZj$ production, and therefore can be emitted from either the initial or final state or the $W$ propagator whilst the Higgs boson only couples to the top quark and $W$ boson, so it cannot be emitted from the initial state.\footnote{As already mentioned, even in the 4FS where $\mb\neq0$ the emission of a Higgs boson from the bottom-quark fermion line is negligible for this process.} This leads to different  kinematics and enhances the differences between the two schemes.

\subsubsection{NLO QCD+EW predictions}

In this section we study differential distributions and explore the 
impact of EW corrections at the differential level. No $t$-channel selection is applied and therefore all the  $t$-, $s$- and $tW_h$-channel contributions are taken into account. Each one of the Figs.~\ref{fig:tHj_EW}--\ref{fig:toplplmj_nocuts_EW} have the same layout, which we describe in the following. In each figure we display four different plots for the following distributions: the pseudorapidity and the transverse momentum of the hardest light-jet, the transverse momentum of the top and of the Higgs/$Z$ boson or $\ell^+ \ell^-$ pair. In each plot we show in the main panel the predictions at different accuracies: $\LO$, $\NLOQCD$ and $\NLOQCDEW$, which is our best prediction. In the first inset we show the theory uncertainty band for the $\NLOQCDEW$ prediction, normalised to its central value. The band is given by the sum in quadrature of scale and PDF uncertainties. We remind the reader that we combine 4FS and 5FS scale uncertainties into the 5FS$_{\rm 4-5}^{\rm scale}$ one, which has already been described in Sec.~\ref{sec:inclusive} and in more detail in Secs.~\ref{sec:flavscheme} and \ref{sec:ch-sepa}. In the second inset we show the scale uncertainty band for the $\NLOQCD$ prediction normalised to its central value. The scale uncertainty band is shown both for the 5FS$_{\rm 4-5}^{\rm scale}$, which is by definition equal to the one of the $\NLOQCDEW$ prediction, and for the 5FS. Also, we show the $\NLOQCDEW/\NLOQCD$ ratio for the central values. One can judge the impact of NLO EW corrections by comparing this ratio with the scale uncertainty in the 5FS$_{\rm 4-5}^{\rm scale}$, and also appreciate the difference with the pure 5FS uncertainty.

We start by commenting on Fig.~\ref{fig:tHj_EW}, where we show the distributions for the $tHj$ process. We have focussed on the observables for which EW corrections are neither negligible, nor flat. For instance, the shapes of the rapidity of the top quark and Higgs boson are not modified by the NLO EW corrections, with the $\NLOQCDEW/\NLOQCD$ ratio being flat over all rapidities and equal to the one already shown in Tab.~\ref{tab:K_factors}.  Similarly, EW corrections to the $b$-jet distributions are negligible. 
NLO EW corrections in general reduce the $tHj$ cross section, in particular in the tails of the transverse momentum distributions. This is 
the typical  behaviour of EW corrections. Only in the central region of the $\eta(j_{l,1})$ distribution we observe a positive effect induced by NLO EW corrections.  The same effect has been observed for single top production in Ref.~\cite{Frederix:2019ubd} and found to be related to the $tW_h$ channel contribution, which enters only at $\NLOQCD$ accuracy and populates the central region of the $\eta(j_{l,1})$ distributions.  Indeed, when a light jet emerges from the $W$-boson decay, no enhancement is present in the region close to the beam-pipe axis, at variance with the light jet emerging from $t$-channel production. This effect can be clearly seen by comparing the LO and $\NLOQCD$ lines in the main panel. For more details on this effect see Appendix A in  Ref.~\cite{Frederix:2019ubd}.

For all distributions, the impact of NLO EW corrections remains within the scale uncertainty 
band of the $\NLOQCD$ results, approaching the lower edge of the band in the tails of the distributions. However, this is true only because we employed the  5FS$_{\rm 4-5}^{\rm scale}$. If we had considered only the 5FS scale uncertainties, this 
would not be the case;  NLO EW corrections would shift the central value of the prediction to the lower edge of the $\NLOQCD$  scale uncertainty band, and outside of it for the $p_T(j_{l,1})$ and $p_T(H)$ distributions.

\begin{figure}[h]
\centering
\includegraphics[width=0.315\textwidth]{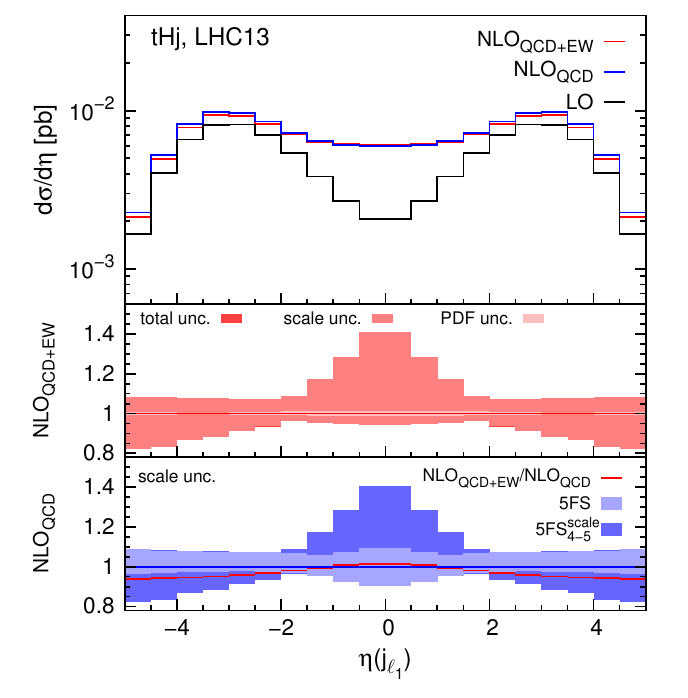}
\includegraphics[width=0.315\textwidth]{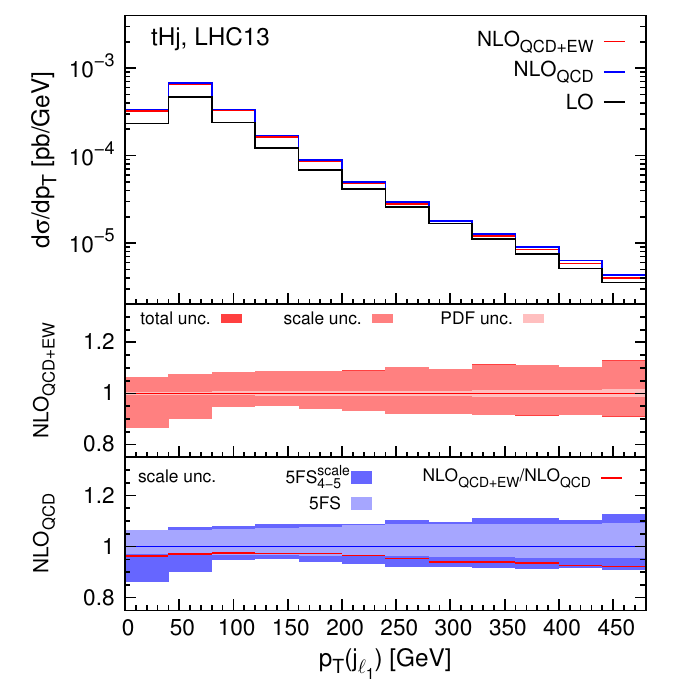}\\
\includegraphics[width=0.315\textwidth]{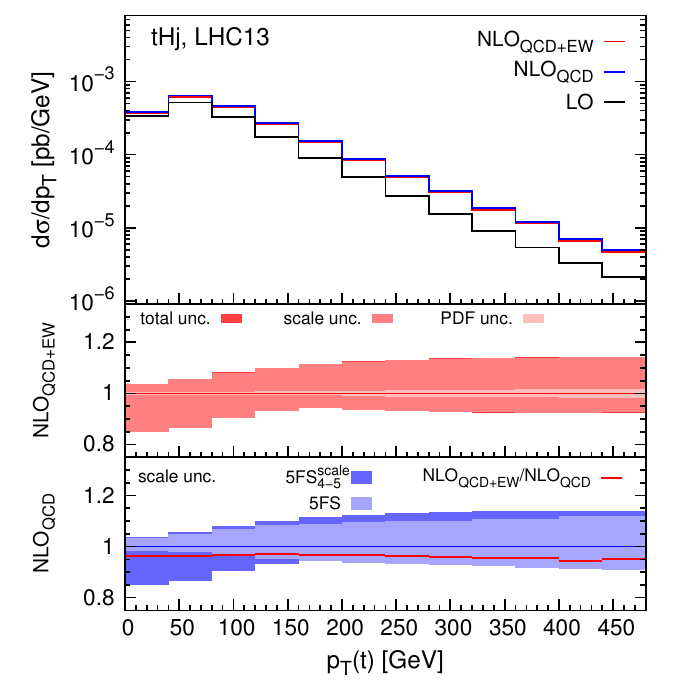}
\includegraphics[width=0.315\textwidth]{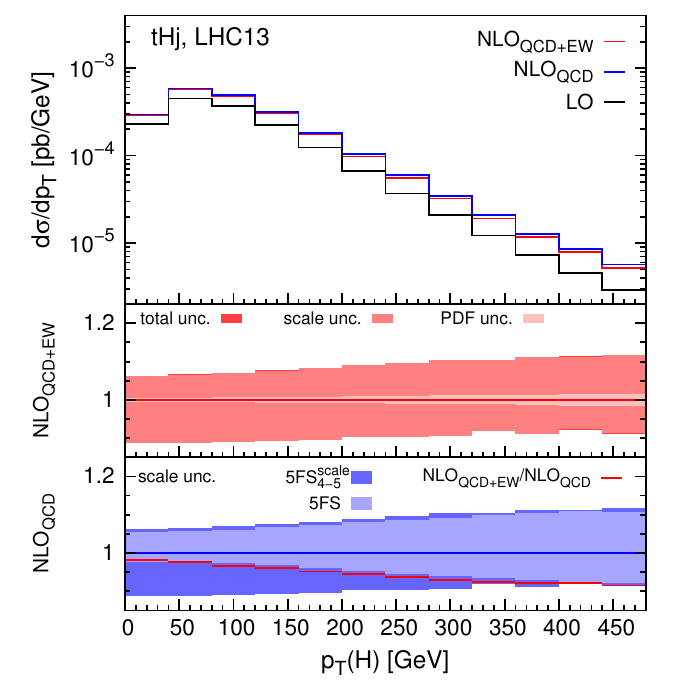}
\caption{$\NLOQCDEW$ predictions for $tHj$. In each plot the first inset shows the total uncertainty (flavour-scheme, scale and PDFs) and the second inset shows the $\NLOQCDEW/\NLOQCD$ ratio along with the $\NLOQCD$ scale uncertainties both in the 5FS$_{\rm 4-5}^{\rm scale}$  and 5FS. } 
\label{fig:tHj_EW}
\end{figure}
\begin{figure}[H]
\centering
\includegraphics[width=0.315\textwidth]{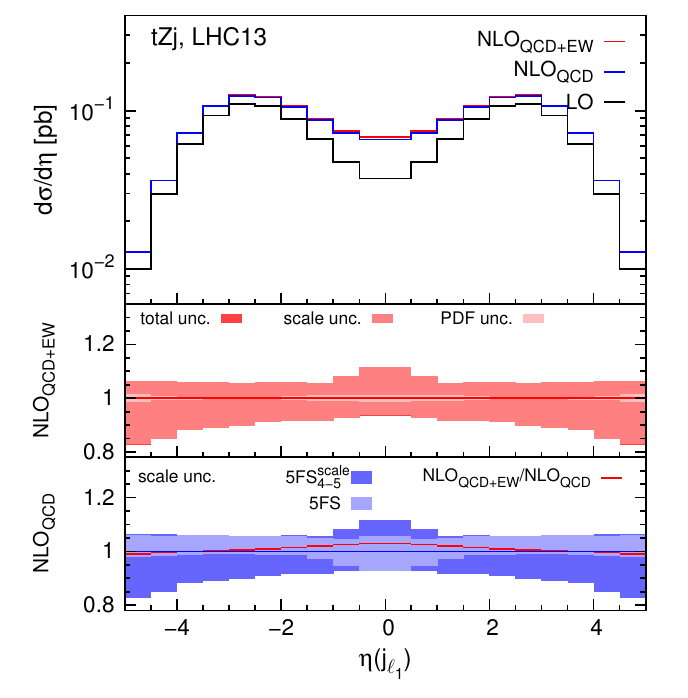}
\includegraphics[width=0.315\textwidth]{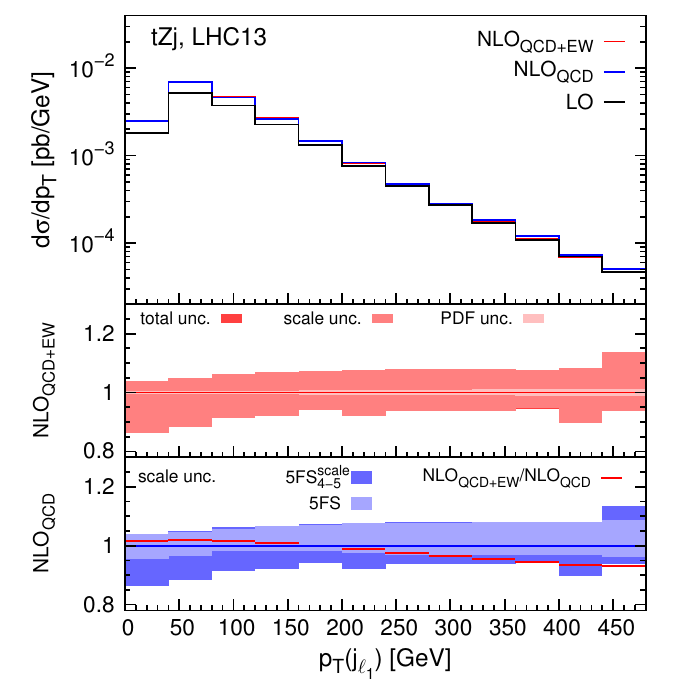}\\
\includegraphics[width=0.315\textwidth]{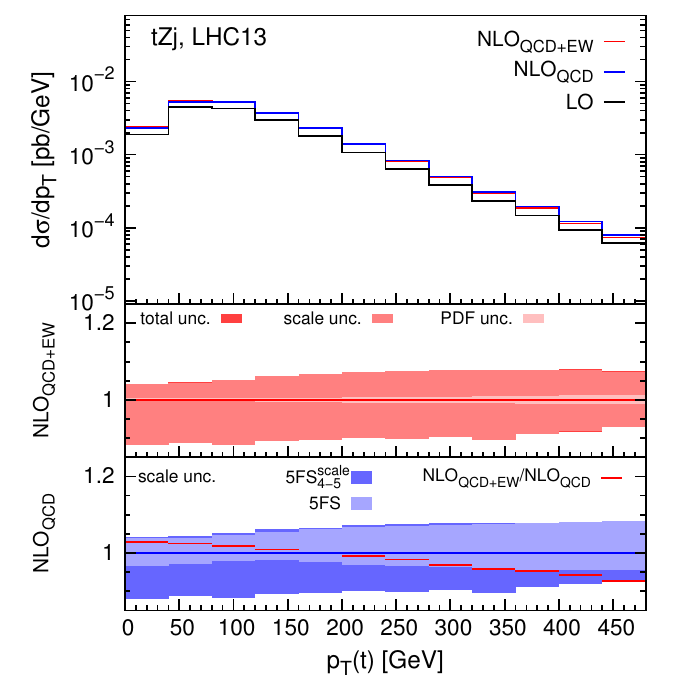}
\includegraphics[width=0.315\textwidth]{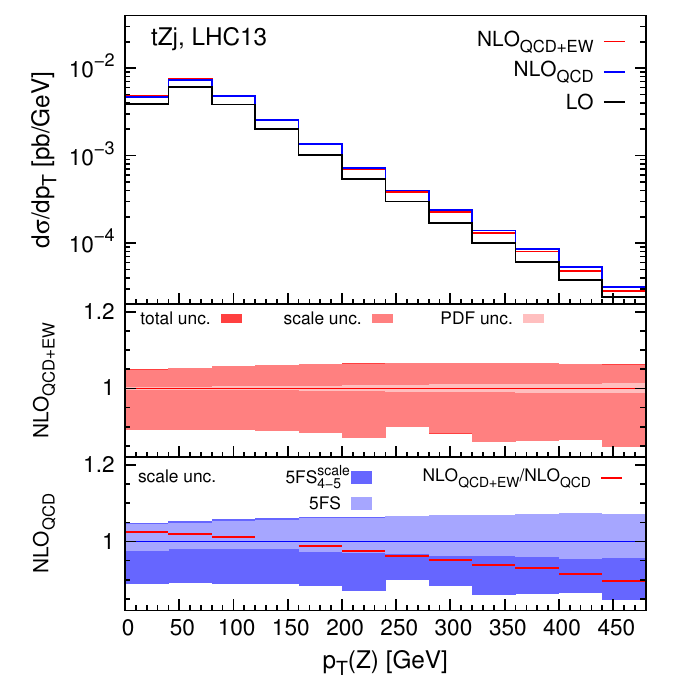}
\caption{$\NLOQCDEW$ predictions for $tZj$. The layout of the plots is the same of Fig.~\ref{fig:tHj_EW}.} 
\label{fig:tZj_EW}
\end{figure}

\begin{figure}[h]
\centering
\includegraphics[width=0.315\textwidth]{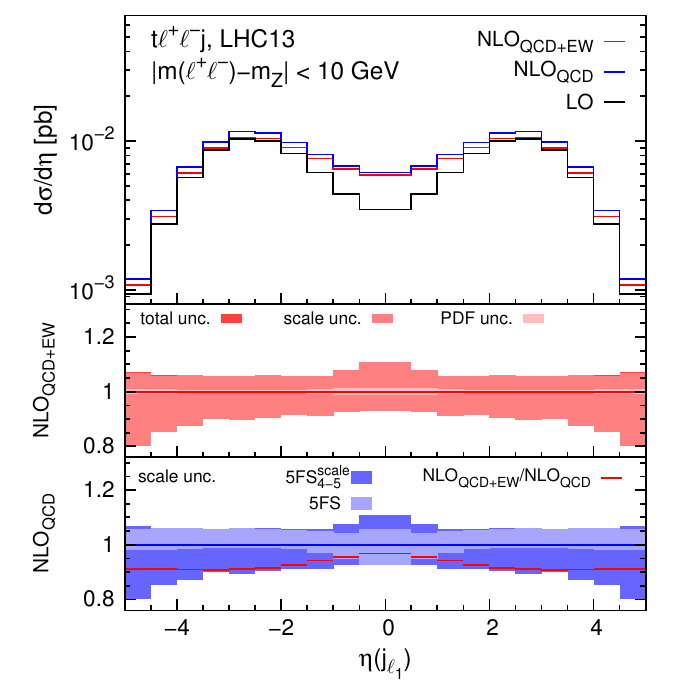}
\includegraphics[width=0.315\textwidth]{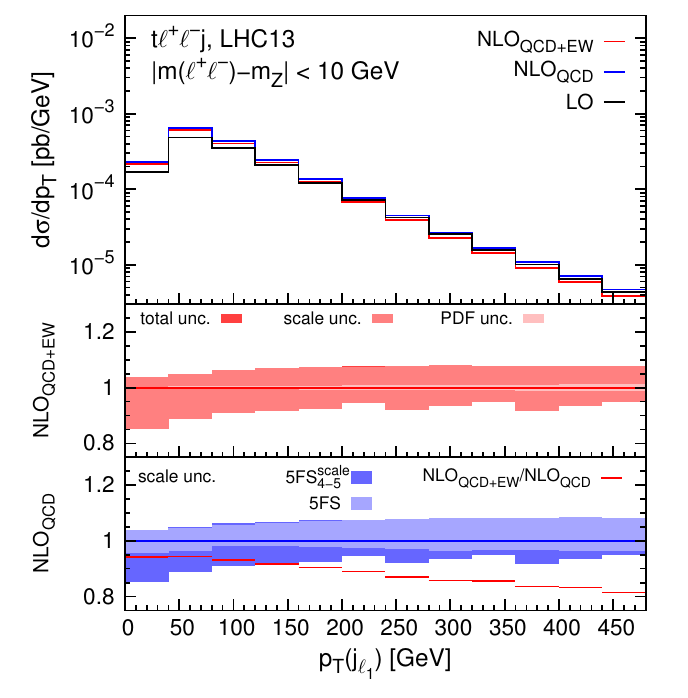}\\
\includegraphics[width=0.315\textwidth]{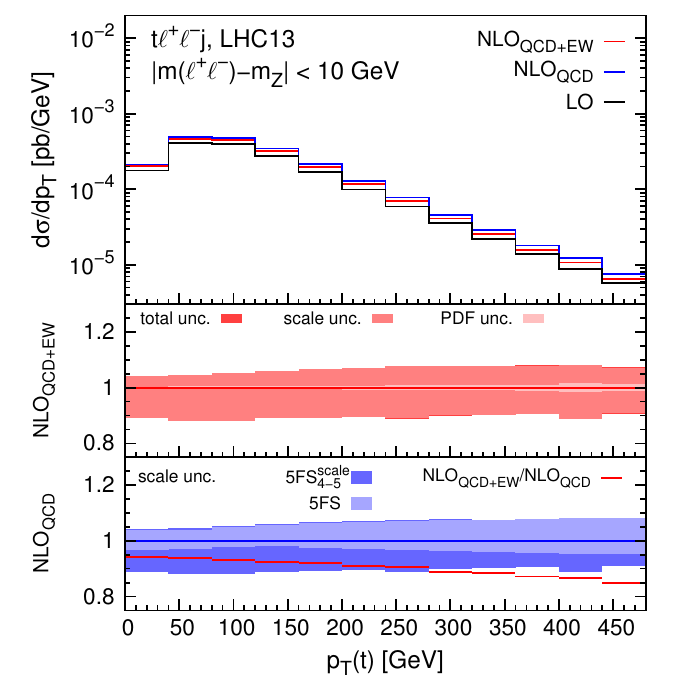}
\includegraphics[width=0.315\textwidth]{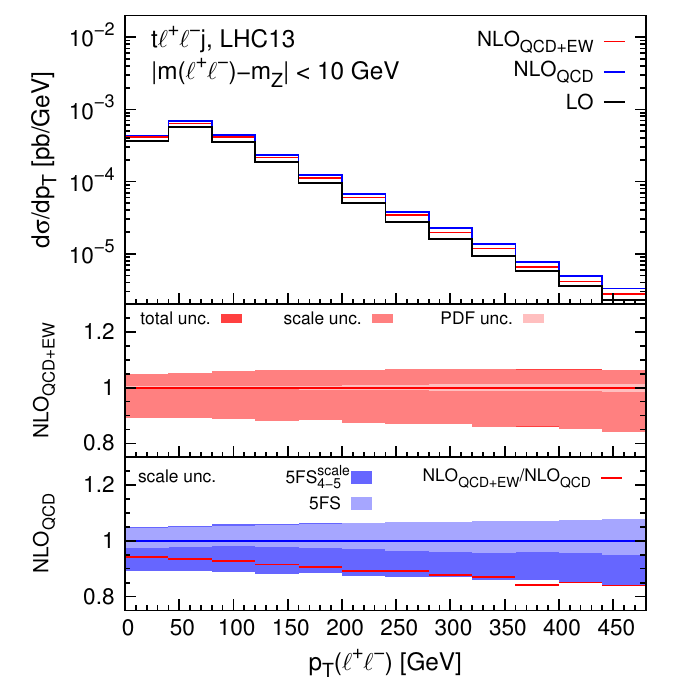}
\caption{$\NLOQCDEW$ predictions for $t\ell^+\ell^-j$  ($Z$-peak). The layout of the plots is the same of Fig.~\ref{fig:tHj_EW}.} 
\label{fig:toplplmj_Zpeak_EW}
\end{figure}
\begin{figure}[H]
\centering
\includegraphics[width=0.315\textwidth]{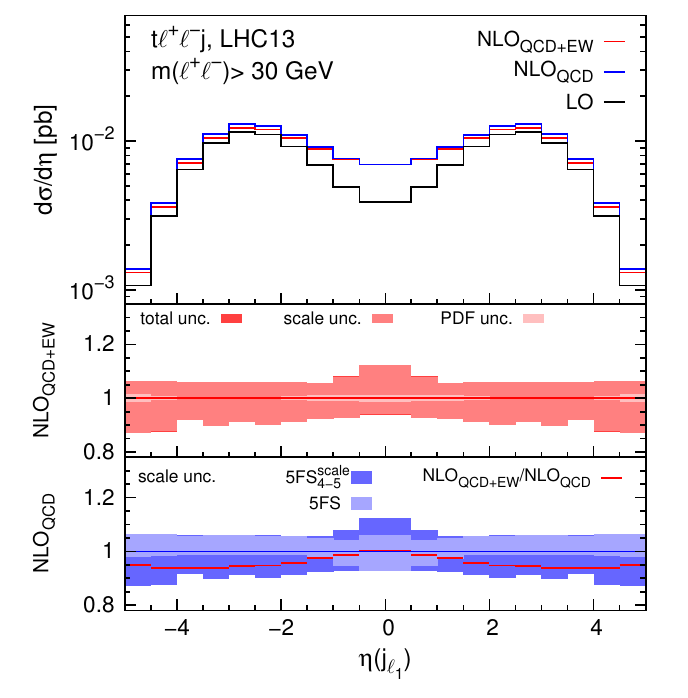}
\includegraphics[width=0.315\textwidth]{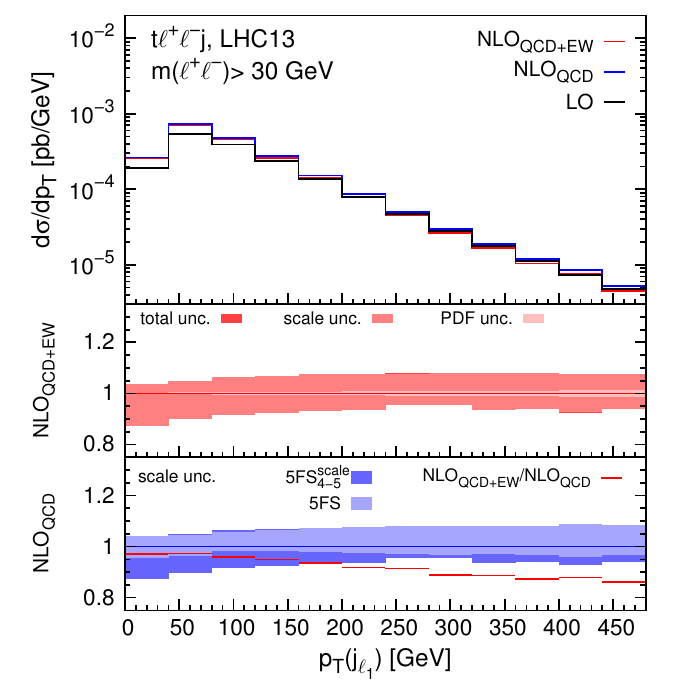}\\
\includegraphics[width=0.315\textwidth]{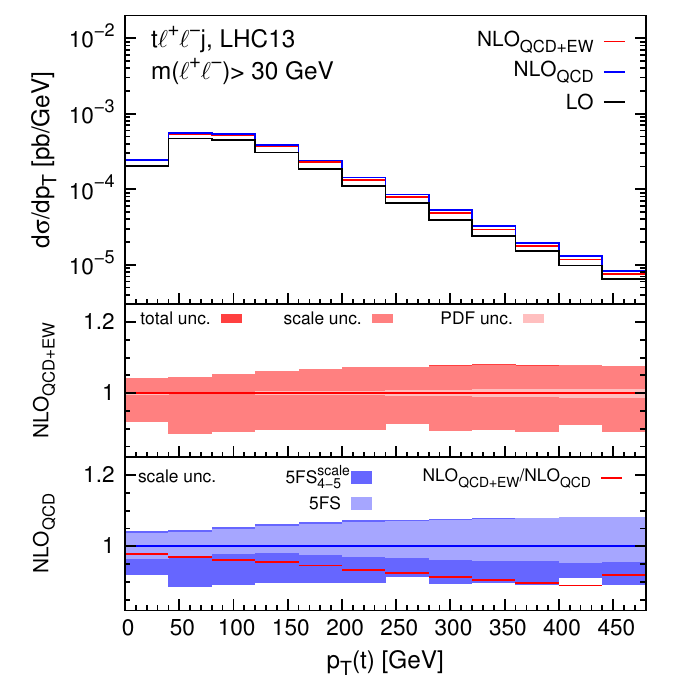}
\includegraphics[width=0.315\textwidth]{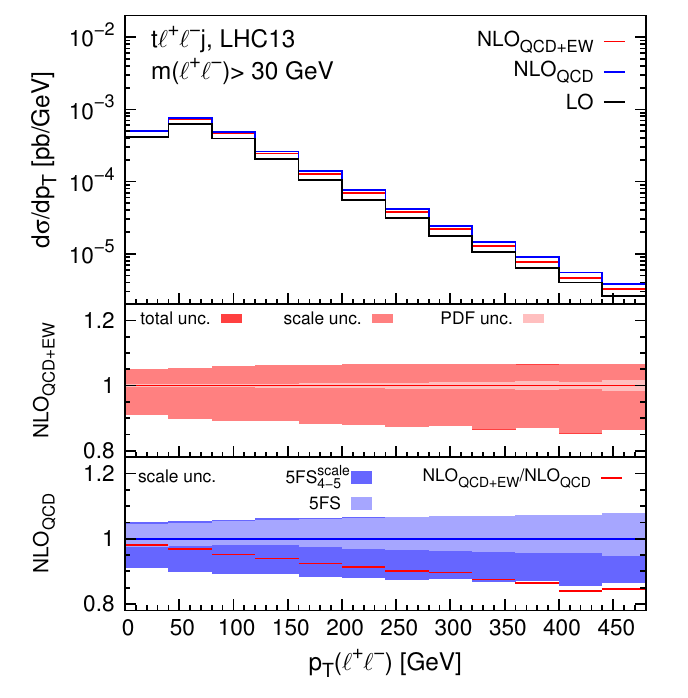}
\caption{$\NLOQCDEW$ predictions for $t\ell^+\ell^-j$  (``inclusive''). The layout of the plots is the same of Fig.~\ref{fig:tHj_EW}.} 
\label{fig:toplplmj_nocuts_EW}
\end{figure}

The corresponding results for $tZj$ production are shown in Fig.~\ref{fig:tZj_EW}. NLO EW corrections in $tZj$ show the same qualitative features as in $tHj$ with the corrections reaching $\sim$10\% in the tail. At low transverse momentum they in fact increase the cross section by a couple of percent. 

\begin{figure}[t!]
\centering
\includegraphics[width=0.315\textwidth]{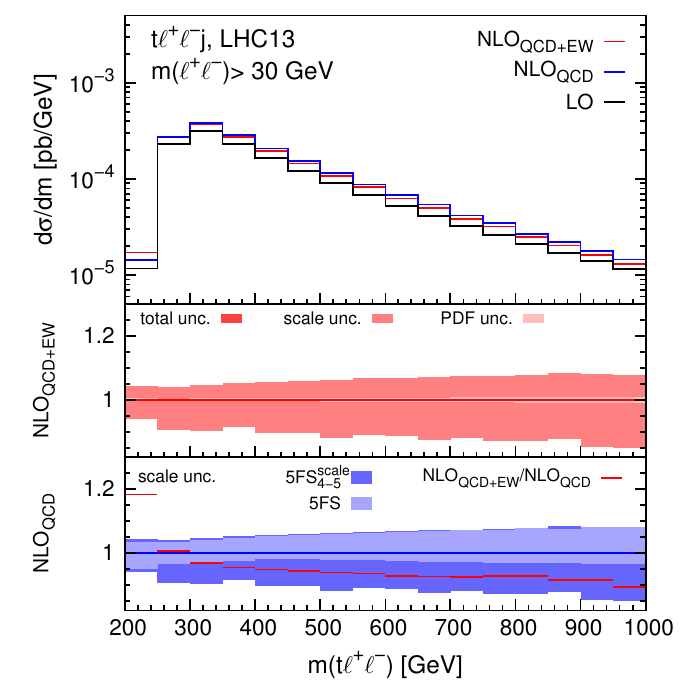}
\includegraphics[width=0.315\textwidth]{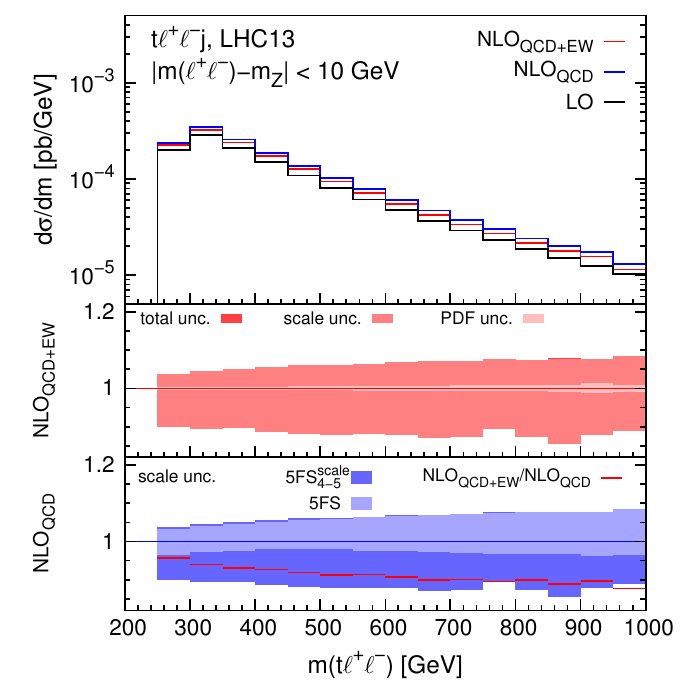}\\
\includegraphics[width=0.315\textwidth]{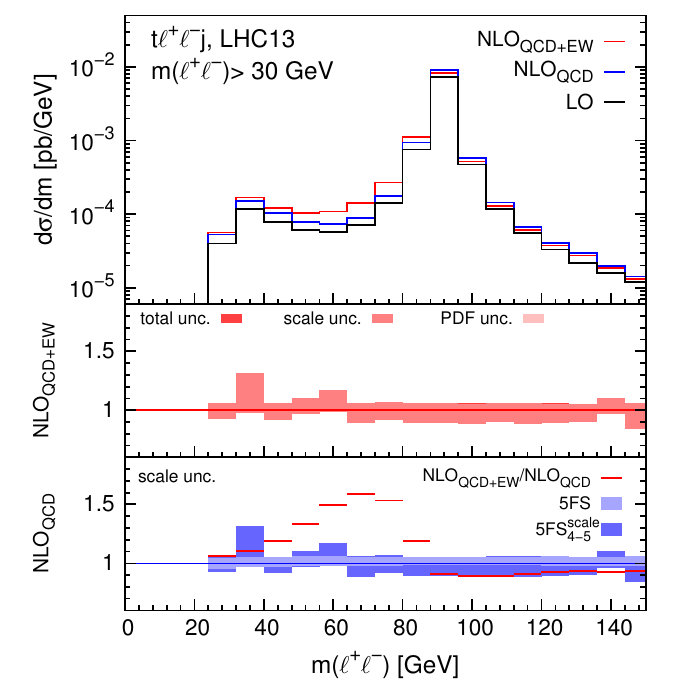}
\includegraphics[width=0.315\textwidth]{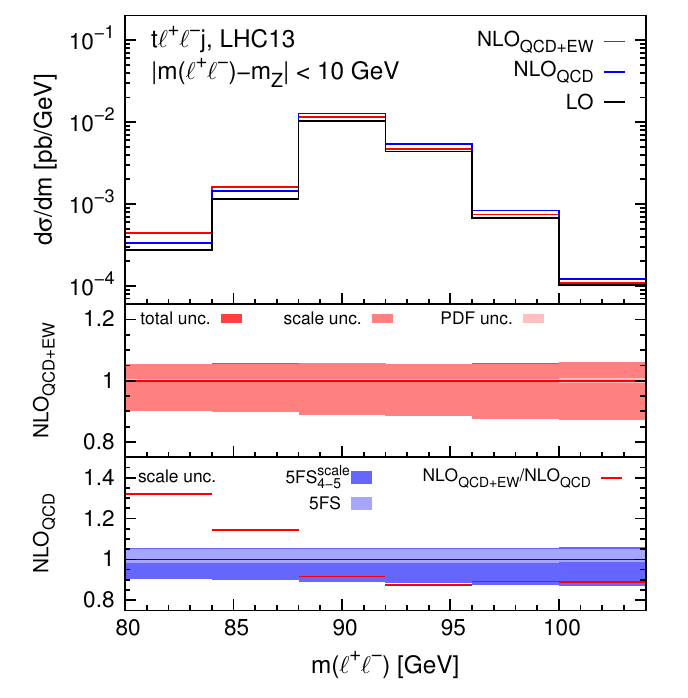}\\
\includegraphics[width=0.315\textwidth]{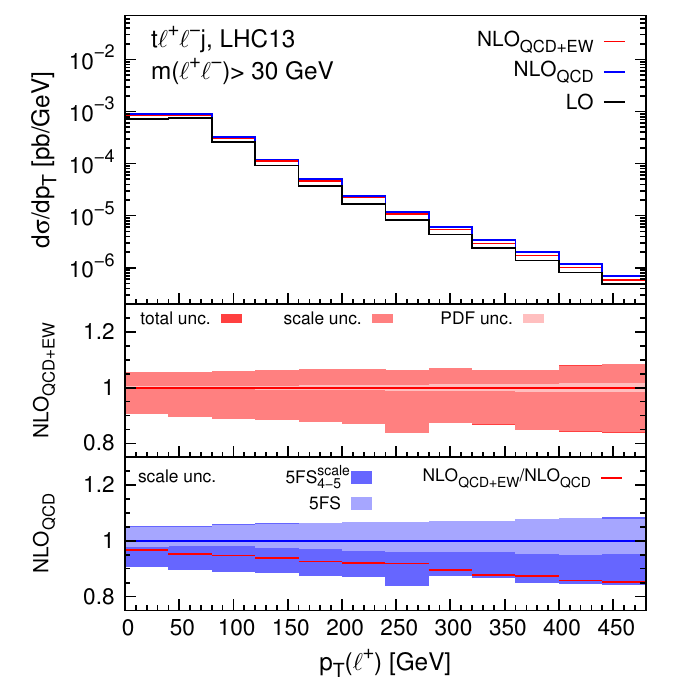}
\includegraphics[width=0.315\textwidth]{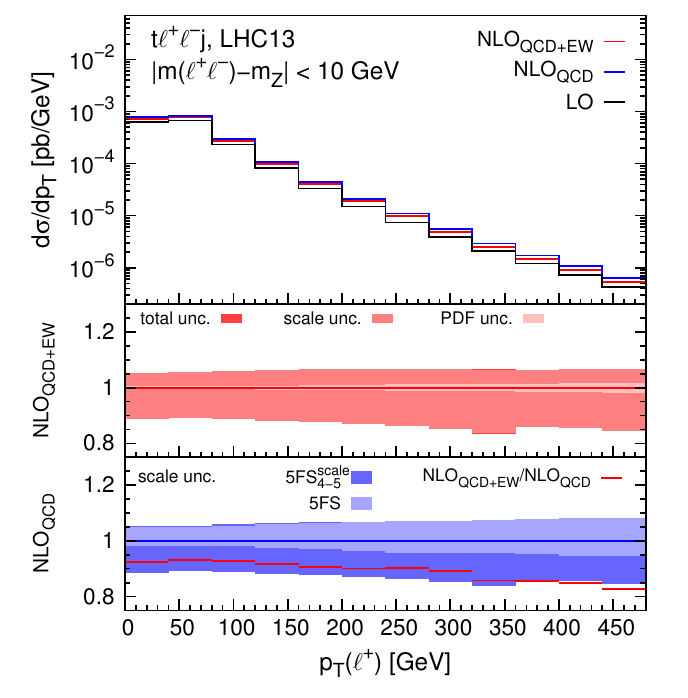}\\
\includegraphics[width=0.315\textwidth]{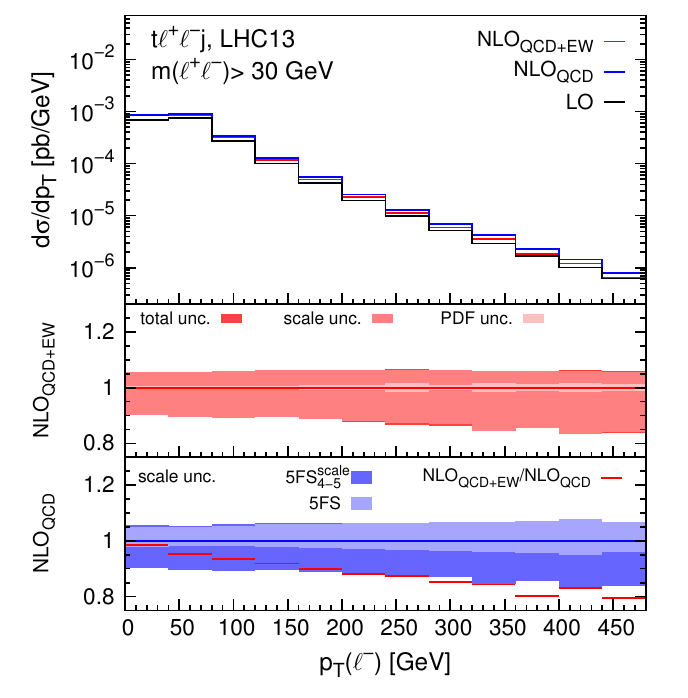}
\includegraphics[width=0.315\textwidth]{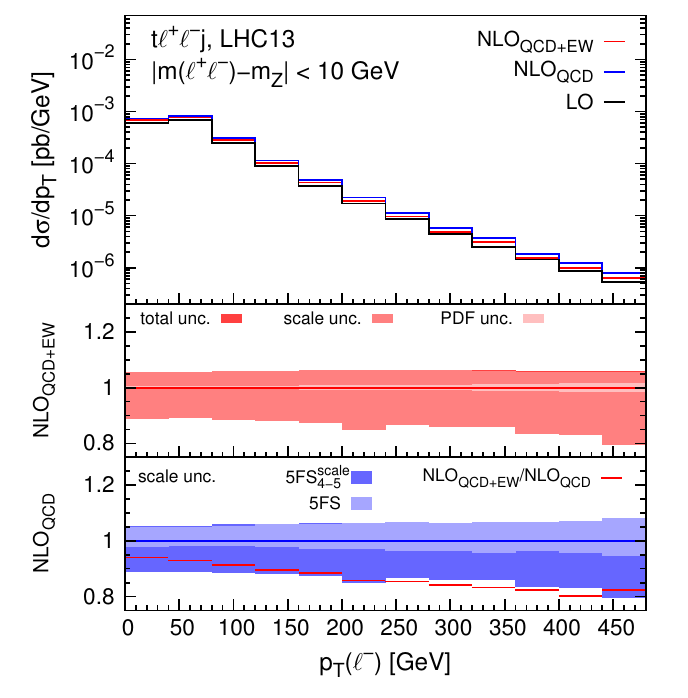}
\caption{Further plots as in Figs.~\ref{fig:toplplmj_Zpeak_EW} and \ref{fig:toplplmj_nocuts_EW} for $t\ell^+\ell^-j$ lepton-based observables. The ``inclusive'' case is displayed on the  left while the  $Z$-peak one on the right.} 
\label{fig:toplplmj_EW}
\end{figure}

The corresponding results for $\tllj$ production are shown in Figs.~\ref{fig:toplplmj_Zpeak_EW} and \ref{fig:toplplmj_nocuts_EW} for the $Z$-peak and  ``inclusive''  selection cuts. Whilst the qualitative behaviour of the NLO EW corrections remains the same as for $tZj$ production, the size of the corrections is larger, reaching up to 15\% in the tails of the distributions. Especially, in contrast with $tZj$ and $tHj$ production, the NLO EW corrections are large enough to possibly lie outside the $\NLOQCD$ scale-uncertainty band also in the 5FS$_{\rm 4-5}^{\rm scale}$. This happens above 150 GeV in the transverse momentum distribution of the light jet and in the last few bins of the top-quark and dilepton transverse momentum distributions.
We notice that in the tails NLO EW corrections are similar in size for $Z$-peak and ``inclusive'' selection cuts, while in the rest of the spectrum they are larger in the latter, consistently with the results in Tab.~\ref{tab:K_factors}. In fact, in the pure 5FS, for the $Z$-peak selection cuts NLO EW corrections would be outside the scale uncertainty band over the full spectrum, with the exception of  central region of the $\eta(j_{l,1})$ distribution.

Finally in the cases of $t\ell^+\ell^-j$, both  $Z$-peak and ``inclusive'', we show additional distributions involving the leptons. In Fig.~\ref{fig:toplplmj_EW} we show the invariant mass of the  $t\ell^+\ell^-$ system, the invariant mass of the lepton pair and the 
transverse momentum distributions of the two leptons. The $t\ell^+\ell^-$ system invariant mass behaves in a similar way for both
 the ``inclusive'' and $Z$-peak results, reaching 10\% reduction in the rate in the tail of the distribution. NLO EW corrections have a large 
 impact on the shape of the dilepton invariant mass distribution. Whilst close to  the $Z$-peak, where the bulk of the cross section originates, corrections are  negative and relatively small,  at lower invariant masses they become very large. Indeed, we see a pronounced bump in the $\NLOQCDEW/\NLOQCD$ ratio below the $Z$ mass. This is related to the photon emission from the leptons, which reduces the lepton invariant mass
  due to events migrating to bins with a smaller invariant mass. The NLO EW corrections increase the rate in the region of $50~ {\rm GeV}<m(\ell^+ \ell^-)< 80 ~ {\rm GeV}$ by up to 60\% compared to the $\NLOQCD$ result.  The same pattern, although smaller in size, is observed even in the 
  $Z$-peak range, with the impact of NLO EW corrections reaching 30\% at the lower end of the distribution. In both cases, the impact of NLO EW corrections is much larger than the $\NLOQCD$ scale-uncertainty band, even in the 5FS$_{\rm 4-5}^{\rm scale}$. Finally, we comment on the lepton $p_T$ distributions. These exhibit the typical behaviour of EW corrections, with large (reaching 20\%) negative corrections in the tails of the distributions. Using the 5FS$_{\rm 4-5}^{\rm scale}$, the NLO EW corrections lie at the lower edge of the QCD scale-uncertainty bands. In the case of a pure 5FS, they would be outside, both in the $Z$-peak and ``inclusive'' case.

\subsection{QCD and QED shower effects}
\label{sec:shower}
In Sec.~\ref{sec:differential} we have computed the NLO EW corrections for various observables and we have found a significant impact in two cases: on the tails of the $p_T$ distributions and especially in the dilepton invariant mass distribution in $\tllj$ production. While the former case is due to purely weak effects, namely Sudakov logarithms, the latter originates from QED final-state radiation (FSR). In this section we therefore
explore the dependence of QED FSR effects on the recombination parameters for leptons and photons. Also, we investigate the impact of the multiple emission of photons via a shower simulation that includes  QED effects. 

To this purpose, we generalise \eqref{eq:recQED} into 
\begin{equation}
\Delta R(\ell, \gamma) < R^{\ell}_{\rm rec} \,, \label{eq:recQED_general}
\end{equation}
and we look at the dependence of cross-section predictions on  the recombination parameter $R^{\ell}_{\rm rec}$.
In the previous sections $R^{\ell}_{\rm rec}$ was set equal to 0.1.  In principle, if no selection cuts were applied on  the leptons, the {\it inclusive} results would not depend on the value $R^{\ell}_{\rm rec}$. However, when we study the $\tllj$ process, in both the ``inclusive'' and the $Z$-peak cases, there is an $m(\ell^+ \ell^-)$ cut applied. Therefore we expect that not only differential distributions but also total rates do depend on  $R^{\ell}_{\rm rec}$.  For this reason, we consider both the $R^{\ell}_{\rm rec}=0.1$ and $R^{\ell}_{\rm rec}=0.5$ options and we determine the impact on the total rates and the lepton-related distributions. 
At the same time, we examine whether NLO EW corrections, when they are dominated by QED FSR, can be equivalently simulated by allowing photon emissions within the QCD shower. For this approach we focus on the $\NLOQCD$ result and we use the default tune of {\sc \small PYTHIA8} \cite{Sjostrand:2007gs, Sjostrand:2014zea}  for the parton shower. In order  to compare these results with the fixed-order ones, we keep the (anti)top quark stable and we switch on the photon emissions from quarks and leptons. Within the analysis, we apply the same lepton-photon recombination ($R^{\ell}_{\rm rec}=0.1,0.5$) and the same jet algorithm as at fixed order.

In Tab.~\ref{tab:recEW} we show the cross sections for  $\tllj$ production. Fixed-order (FO) $\NLOQCDEW$ results are shown for $R^{\ell}_{\rm rec}=0.1$ (the same number of Tab.~\ref{tab:comp_xsec}) and  $R^{\ell}_{\rm rec}=0.5$ and they are compared to the $\NLOQCD$ results, which are shown for different set-ups: fixed-order (FO), matched to the QCD parton shower ($\PSQCD$) via the {\sc MC@NLO} method \cite{Frixione:2002ik}  and including also QED effects in the shower  ($\PSQCDQED$). In the last case, results are again shown for $R^{\ell}_{\rm rec}=0.1,0.5$. All the scale uncertainties are in the standard 5FS.
\begin{table}[t]
\renewcommand{\arraystretch}{2.5}
\scriptsize
\begin{center}
\begin{tabular}{c | c | c c c}
\multicolumn{4}{c}{\normalsize $t \ell^+ \ell^- j$ [fb]} \\
\hline
{\normalsize Order} & settings & ``inclusive'' & $Z$-peak &  \\
\hline
\multirow{2}{*}{$\NLOQCDEW$} & FO, $R^{\ell}_{\rm rec}=0.1$ & $89.6(2)_{-1.7 (-1.9  \%)}^{+5.1 (+5.7  \%)}~_{-0.4 (-0.4  \%)}^{+0.4 (+0.4  \%)}$ & $77.2(2)_{-1.5 (-1.9  \%)}^{+4.9 (+6.3  \%)}~_{-0.3 (-0.4  \%)}^{+0.3 (+0.4  \%)}$ \\
 & FO, $R^{\ell}_{\rm rec}=0.5$ & $89.5(2)_{-1.7 (-1.9  \%)}^{+5.1 (+5.7  \%)}~_{-0.4 (-0.4  \%)}^{+0.4 (+0.4  \%)}$ & $78.1(2)_{-1.5 (-1.9  \%)}^{+4.8 (+6.1  \%)}~_{-0.3 (-0.4  \%)}^{+0.3 (+0.4  \%)}$ \\
\hline
\multirow{4}{*}{$\NLOQCD$} & FO & $93.7(2)_{-1.7 (-1.8  \%)}^{+4.9 (+5.2  \%)}~_{-0.4 (-0.4  \%)}^{+0.4 (+0.4  \%)}$ & $83.4(2)_{-1.5 (-1.8  \%)}^{+4.3 (+5.1  \%)}~_{-0.4 (-0.4  \%)}^{+0.4 (+0.4  \%)}$ \\
 & $\PSQCD$  & $94.0(2)_{-1.7 (-1.8  \%)}^{+4.8 (+5.1  \%)}~_{-0.4 (-0.4  \%)}^{+0.4 (+0.4  \%)}$ & $83.7(2)_{-1.5 (-1.8  \%)}^{+4.3 (+5.1  \%)}~_{-0.4 (-0.4  \%)}^{+0.4 (+0.4  \%)}$ \\
  & $\PSQCDQED$ , $R^{\ell}_{\rm rec}=0.1$ & $93.8(2)_{-1.7 (-1.8  \%)}^{+4.8 (+5.1  \%)}~_{-0.4 (-0.4  \%)}^{+0.4 (+0.4  \%)}$ & $81.2(2)_{-1.5 (-1.8  \%)}^{+4.1 (+5.1  \%)}~_{-0.4 (-0.4  \%)}^{+0.4 (+0.4  \%)}$ \\
  & $\PSQCDQED$ , $R^{\ell}_{\rm rec}=0.5$ & $93.9(2)_{-1.7 (-1.8  \%)}^{+4.8 (+5.1  \%)}~_{-0.4 (-0.4  \%)}^{+0.4 (+0.4  \%)}$ & $82.3(2)_{-1.5 (-1.8  \%)}^{+4.2 (+5.1  \%)}~_{-0.4 (-0.4  \%)}^{+0.4 (+0.4  \%)}$ \\
\end{tabular}
\end{center}
\caption{Cross section comparisons for $t \ell^+ \ell^- j$.}
\label{tab:recEW}  
\end{table}

Concerning the FO results at $\NLOQCDEW$ accuracy, for the  ``inclusive''  result there is no visible difference by varying the $R^{\ell}_{\rm rec}$, whereas for the  $Z$-peak one the cross section is slightly increased with $R^{\ell}_{\rm rec}=0.5$. Indeed, by increasing the $R^{\ell}_{\rm rec}$ value, more photons are recombined with the leptons and consequently the migration of events away from the reconstructed $Z$ peak is reduced. This effect is negligible for  the  ``inclusive'' case because the $m(\ell^+ \ell^-)$ cut is set far below the $Z$ peak. 

Even with the presence of the $m(\ell^+ \ell^-)$ cut,  $\NLOQCD$ results at FO and matched to $\PSQCD$ are compatible within the statistical error, which is at the permille level. Moving to the $\PSQCDQED$ predictions, in the case of  ``inclusive'' results the differences with the $\PSQCD$ case and among different  $R^{\ell}_{\rm rec}$ choices are within the statistical error. On the contrary,  in the case of $Z$-peak results, the cross section slightly reduces once the photon emission is enabled in the shower and it depends on the value of $R^{\ell}_{\rm rec}$; it increases by increasing $R^{\ell}_{\rm rec}$. Still, similarly to the FO case at $\NLOQCDEW$ accuracy, the differences are within 5FS QCD scale uncertainties.

By looking only at total rates obtained with the same $R^{\ell}_{\rm rec}$ value, it is difficult to determine, especially in the $Z$-peak case, the source of the difference between the FO $\NLOQCDEW$ predictions and the $\NLOQCD$ results matched with $\PSQCDQED$. In particular, it is not clear if this difference originates from the multiple emission of photons, which is only present in $\PSQCDQED$, or the purely weak part of the NLO corrections, which is only included in the $\NLOQCDEW$ predictions. In order to better understand this issue, it is instructive to repeat the same comparison at the differential level, in particular for the $m(\ell^+ \ell^-)$ distribution, which is highly sensitive to FSR QED radiation, as shown in Fig.~\ref{fig:toplplmj_EW}.

\begin{figure}[t]
\centering
\includegraphics[width=0.35\textwidth]{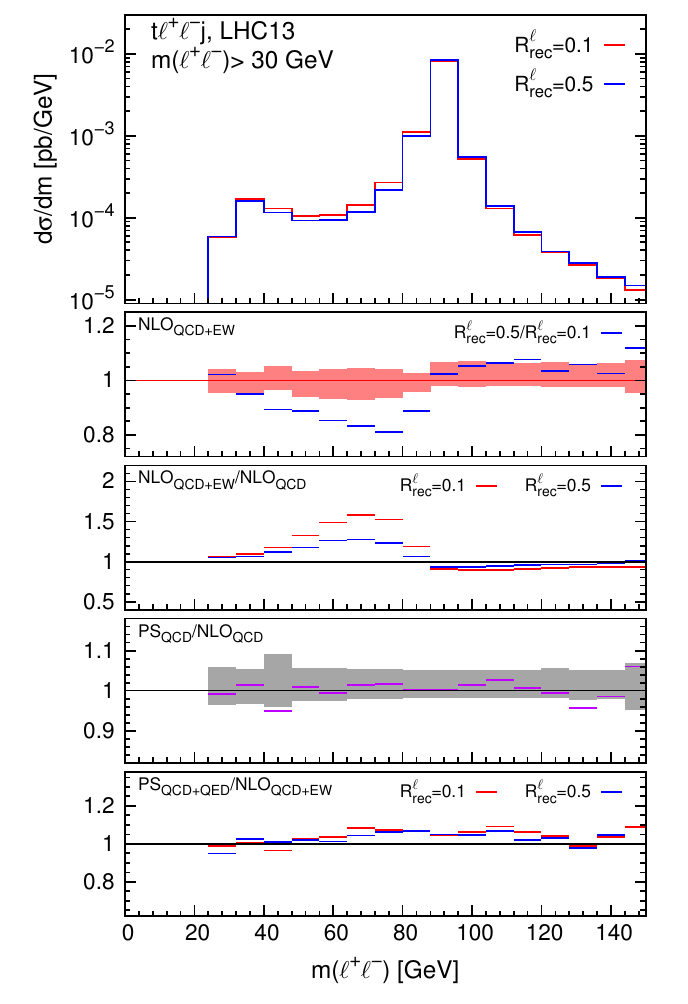}
\includegraphics[width=0.35\textwidth]{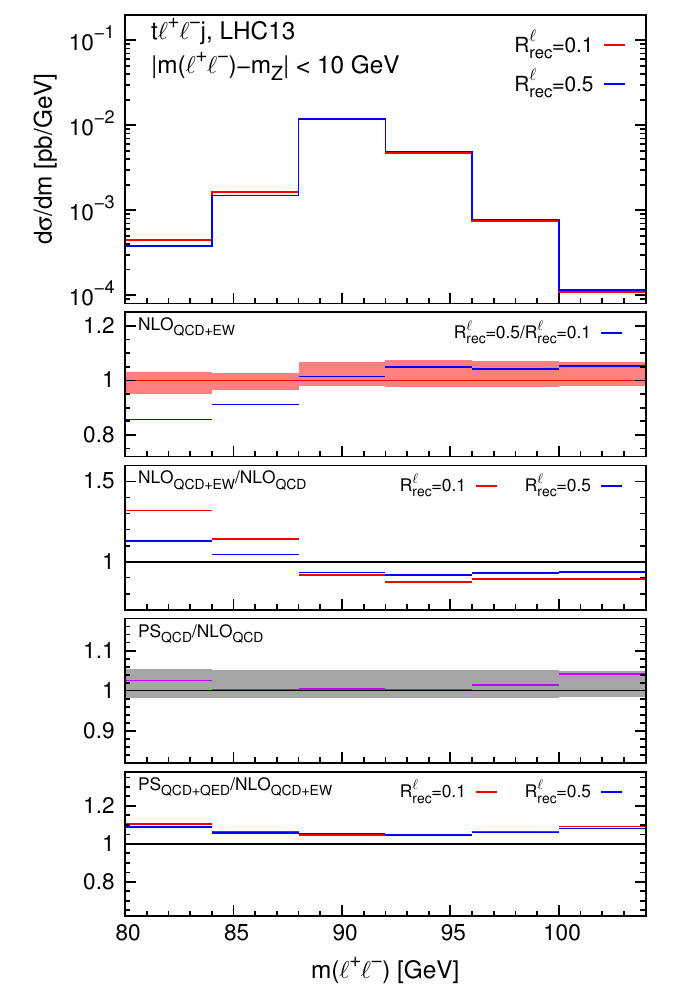} \\
\caption{The $m(\ell^+ \ell^-)$ distribution in $t\ell^+\ell^-j$ production in the ``inclusive'' (left) and  $Z$-peak (right) range for different recombination parameters. See the main text for details. } 
\label{fig:Mvec_FO_PS}
\end{figure}
\noindent
In the plots of Fig.~\ref{fig:Mvec_FO_PS} we compare $m(\ell^+ \ell^-)$ predictions for  different $R^{\ell}_{\rm rec}$ values, for the ``inclusive'' case (left) and the $Z$-peak one (right). In fact, in this figure, the right plot is a zoomed version and with smaller bins of the one on the left. In the main panel, we show results at FO $\NLOQCDEW$ accuracy, for $R^{\ell}_{\rm rec}=0.5$ and $R^{\ell}_{\rm rec}=0.1$. In the first inset we show the ratio of these two different results, together with scale uncertainties for the latter, again in the 5FS. In the second inset we show the $\NLOQCDEW/\NLOQCD$ ratio, for both $R^{\ell}_{\rm rec}$ values. In the third and fourth inset, shower effects are compared to the fixed-order calculation. In particular, in the third, we show the ratio of $\NLOQCD$ predictions matched with $\PSQCD$ and at FO, which does not depend on $R^{\ell}_{\rm rec}$, while in the fourth we show the ratio of $\NLOQCD$ predictions matched with $\PSQCDQED$ and $\NLOQCDEW$ at FO, again for both $R^{\ell}_{\rm rec}$ values.

The first comment on plots of  Fig.~\ref{fig:Mvec_FO_PS} is that the migration of events to lower $m(\ell^+ \ell^-)$ values depends on the $R^{\ell}_{\rm rec}$ value, with a much smaller migration for $R^{\ell}_{\rm rec}=0.5$, as can be seen in the first and second insets. However, this dependence is almost identical in the case of  FO $\NLOQCDEW$ predictions or $\NLOQCD$ ones matched with a $\PSQCDQED$ shower. In fact, the results obtained with these two different simulations are very similar in shape (differences are at the 5--10\% level for the normalisation) also close to the $Z$ resonance, as can be seen in the fourth inset. These differences are mainly induced by  electroweak effects, not the QCD ones. Indeed, as can be seen in the fourth insets,  the differences between the two aforementioned approximations are larger than the differences between $\NLOQCD$ predictions at FO and matched with $\PSQCD$.  Thus, the differences between FO $\NLOQCDEW$ predictions and those at $\NLOQCD$ accuracy matched with a $\PSQCDQED$ shower that are observed in the fourth insets can originate only from two effects: either the purely weak part of the NLO EW corrections or the emissions of photons beyond the first one, which are not part of the FO NLO EW corrections. Since the ratios in the fourth insets are flat and especially do not depend on the value of $R^{\ell}_{\rm rec}$, the differences between the two approximations have to be mainly induced by purely weak contributions from the FO NLO EW corrections. 

Summarising, a shower simulation including QED effects, $\PSQCDQED$,  captures very well the effects from NLO EW corrections for the $m(\ell^+ \ell^-)$ distribution, within a 5--10\% level. This difference is quite flat and is mainly induced by purely weak effects at fixed order. This fact has two consequences. First, also in the case of total rates in Tab.~\ref{tab:recEW} we can safely conclude that the differences observed between $\NLOQCDEW$ and $\PSQCDQED$ results has this origin. Second, by performing a proper matching of the FO $\NLOQCDEW$ calculation and $\PSQCDQED$ simulations, one expects to find  a negligible difference  w.r.t. the pure FO $\NLOQCDEW$ result.

We want to stress that however the $\PSQCDQED$ parton shower cannot  in general capture the impact of NLO EW corrections, {\it e.g.},  in boosted regimes the  purely weak corrections can be large and negative.
\begin{figure}[t]
\centering
 \includegraphics[width=0.35\textwidth]{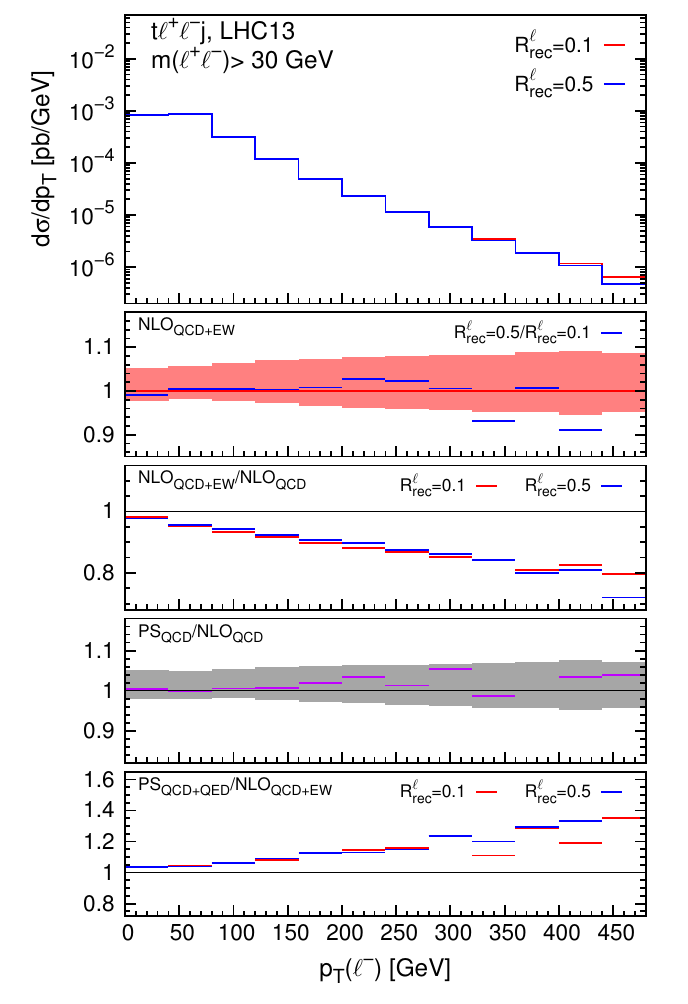}
 \includegraphics[width=0.35\textwidth]{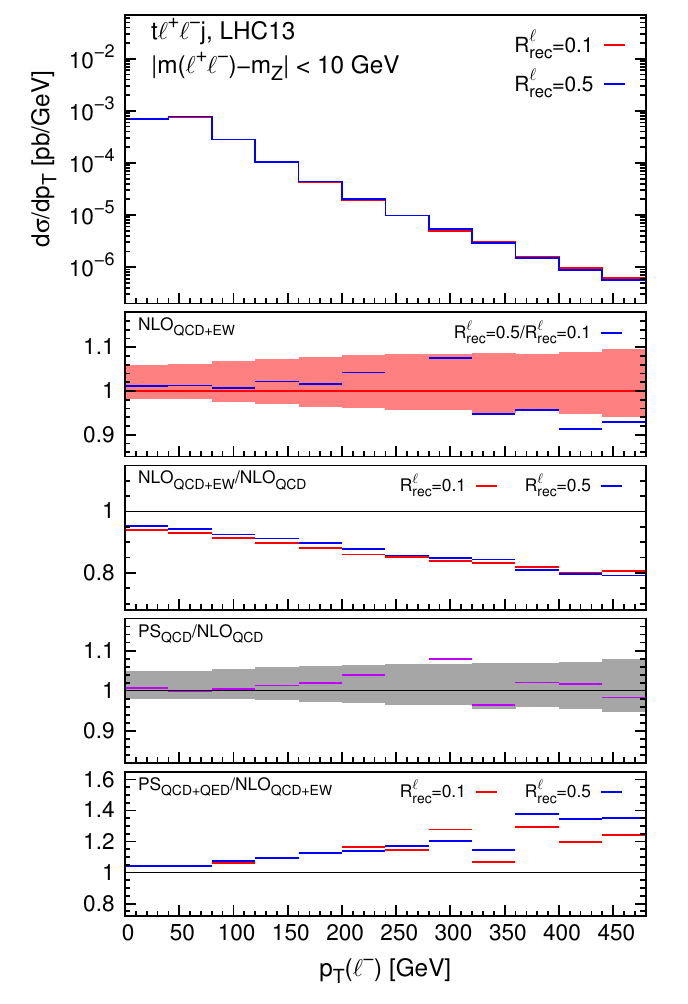} \\
\caption{The $p_T(\ell^-)$ distribution in $t\ell^+\ell^-j$ production in the ``inclusive'' (left) and  $Z$-peak (right) range for different recombination parameters. The layout of the plots is the same of Fig.~\ref{fig:Mvec_FO_PS}.} 
\label{fig:pTlp_PS}
\end{figure}
\noindent
As an evidence of this behaviour we show in Fig.~\ref{fig:pTlp_PS} two plots for the $p_T(\ell^-)$ distribution, using the same layout of those of Fig.~\ref{fig:Mvec_FO_PS}. At variance with the $m(\ell^+ \ell^-)$ distribution, results are almost insensitive to the $R^{\ell}_{\rm rec}$ value, but also in this case the inclusion of the QED shower does not have a sizeable effect. The relevance of purely weak corrections can be seen in the last inset. In the tail of the distribution their impact is large and negative and the $\NLOQCD$  simulation matched with the $\PSQCDQED$ cannot capture their effect.

In conclusion, whilst  a matched simulation at $\NLOQCDEW$ accuracy with $\PSQCDQED$ would further improve the precision, it is not urgently needed for the foreseen accuracy that can be achieved in the next  measurements of $\tllj$ production, and even more $tHj$ production, at the LHC. We leave this possible improvement to future works.

 \section{Conclusions}\label{sec:conclusion}

In this paper we have presented and thoroughly discussed  the calculation of NLO QCD and EW corrections to the production cross section of a single top (anti)quark in association with either a Higgs ($tHj$) or a $Z$ boson ($tZj$) at the LHC. In the context of $tZj$ production, the more realistic $\tllj$ final state has also been considered, taking into account off-shell effects and diagrams where the $\ell^+\ell^-$ pair emerges from a photon propagator. The calculation has been performed in the 5FS via the public version of the code {\sc \small MadGraph5\_aMC@NLO} and we have carefully analysed the comparison with predictions obtained in the 4FS in order to estimate the uncertainty due to the flavour-scheme choice. In our calculation, in order to be closer to the experimental measurements, we do not select a specific production mode ($t$-channel, $s$-channel or $tW$ associated production with subsequent hadronic $W$-boson decays). Moreover, the separation of the different production modes is not properly defined at NLO EW accuracy and in general at higher orders. For this reason, the comparison of 4FS and 5FS predictions and in turn the estimation of the flavour-scheme uncertainty is not trivial. To this purpose,  we have devised and motivated in detail a strategy, denoted in the text as 5FS$_{\rm 4-5}^{\rm scale}$, where the central value is the one given by the 5FS prediction, either at $\NLOQCD$ or $\NLOQCDEW$ accuracy, while the relative scale+flavour-scheme uncertainty band is given by the envelope of the 5FS and 4FS scale uncertainties of $\NLOQCD$ predictions from $t$-channel only contributions. Our best predictions, namely at $\NLOQCDEW$ accuracy in the 5FS$_{\rm 4-5}^{\rm scale}$, for the LHC at the collision energy of 13 TeV are
\begin{align}
\sigma(tHj) &=  82.2 ~{\rm fb} ~_{-10.9  \%}^{+\phantom{0}7.2  \%}~({\rm scale+flavour})~_{-0.6  \%}^{+0.6  \%} ~({\rm PDFs}) \, ,\nonumber \\
\sigma(tZj) &=  \phantom{.}904~{\rm fb} ~_{-11.1  \%}^{+\phantom{0}5.5  \%}~({\rm scale+flavour})~_{-0.4  \%}^{+0.4  \%}~({\rm PDFs})\, , \nonumber \\
\sigma(\tllj) &=  89.6~{\rm fb} ~_{ -10.4  \%}^{+\phantom{0}5.6  \%}~({\rm scale+flavour})~_{-0.4  \%}^{+0.4  \%}~({\rm PDFs})~~~~~{\rm for}~~m(\ell^+ \ell^-)>30~{\rm GeV}\, ,\nonumber \\
\sigma(\tllj) &= 77.2 ~{\rm fb}~_{-11.3  \%}^{+\phantom{0}5.6  \%}~({\rm scale+flavour})~_{-0.4  \%}^{+0.4  \%}~({\rm PDFs})~~~~~{\rm for}~~|m(\ell^+ \ell^-)-m_Z|<10~{\rm GeV} \, ,\nonumber
\end{align}
where each cross section refers to the sum of the case of a top quark and a top antiquark.

 The size of the EW corrections is for all four cases smaller than the scale+flavour uncertainties, which is purely of QCD origin.  However, if we had considered the 5FS only, they would have been (much) larger than the scale uncertainties, with the exception of $\sigma(tZj)$. A similar pattern has been observed also for differential distributions. On the other hand, for large transverse momenta ($\sim300$ GeV) of the light jet or the heavy boson, the EW corrections are as large as ($tHj$ and $tZj$) or even larger ($\tllj$, especially by requiring $|m(\ell^+ \ell^-)-m_Z|<10~{\rm GeV}$) than the scale+flavour uncertainties in the 5FS$_{\rm 4-5}^{\rm scale}$. 
 
Finally, in the case of $\tllj$ production, we have also compared fixed-order predictions at $\NLOQCDEW$ accuracy with $\NLOQCD$ predictions matched with $\PSQCDQED$, a parton shower simulation including also multiple  photon emissions. 
First, we have verified that in both approximations total rates are almost insensitive to the photon-lepton recombination parameter. Second, we have shown that the $m(\ell^+ \ell^-)$ spectrum at $\NLOQCDEW$ accuracy can be very well reproduced by the 
$\NLOQCD$ calculation matched with $\PSQCDQED$. Differences are quite flat and at the 5--10\% level and they originate from the purely weak component of NLO EW corrections, which also explains the differences observed for total rates in the two different approximations. On the contrary, the corrections induced by the $\PSQCDQED$ beyond the first photon emission, which is already included in the $\NLOQCDEW$ calculation, are negligible. Finally, we have explicitly shown that for other lepton-based observables, such as the $p_T(\ell^-)$ distributions, the purely weak component of NLO EW corrections can be non-negligible and therefore the $\PSQCDQED$ cannot correctly reproduce the $\NLOQCDEW$ predictions.

 \section*{Acknowledgements}

We want to thank Rikkert Frederix and Marco Zaro for interesting discussions and suggestions. We are grateful to   the developers of {\sc MadGraph5\_aMC@NLO} for the long-standing collaboration and for discussions.    
   The work of D.~P.~is supported by the Deutsche Forschungsgemeinschaft (DFG) under Germany's Excellence Strategy - EXC 2121 ``Quantum Universe'' - 390833306. The work of I.T. is supported by the Swedish Research Council under contract number 2016-05996.

\bibliographystyle{utphys}

 \bibliography{article} 

\providecommand{\href}[2]{#2}\begingroup\raggedright\begin{thebibliography}{10}

\bibitem{Sirunyan:2018zgs}
{\bfseries CMS} Collaboration, A.~M. Sirunyan {\em et~al.}, ``{Observation of
  Single Top Quark Production in Association with a $Z$ Boson in Proton-Proton
  Collisions at $\sqrt {s}$ =13 TeV''},
  \href{http://dx.doi.org/10.1103/PhysRevLett.122.132003}{{\em Phys. Rev.
  Lett.} {\bfseries 122} no.~13, (2019) 132003},
\href{http://arxiv.org/abs/1812.05900}{{\ttfamily arXiv:1812.05900 [hep-ex]}}.

\bibitem{Sirunyan:2017nbr}
{\bfseries CMS} Collaboration, A.~M. Sirunyan {\em et~al.}, ``{Measurement of
  the associated production of a single top quark and a Z boson in pp
  collisions at $\sqrt{s} =$ 13 TeV''},
  \href{http://dx.doi.org/10.1016/j.physletb.2018.02.025}{{\em Phys. Lett.}
  {\bfseries B779} (2018) 358--384},
\href{http://arxiv.org/abs/1712.02825}{{\ttfamily arXiv:1712.02825 [hep-ex]}}.

\bibitem{Aaboud:2017ylb}
{\bfseries ATLAS} Collaboration, M.~Aaboud {\em et~al.}, ``{Measurement of the
  production cross-section of a single top quark in association with a Z boson
  in proton?proton collisions at 13 TeV with the ATLAS detector''},
  \href{http://dx.doi.org/10.1016/j.physletb.2018.03.023}{{\em Phys. Lett.}
  {\bfseries B780} (2018) 557--577},
\href{http://arxiv.org/abs/1710.03659}{{\ttfamily arXiv:1710.03659 [hep-ex]}}.

\bibitem{Aad:2020wog}
{\bfseries ATLAS} Collaboration, G.~Aad {\em et~al.}, ``{Observation of the
  associated production of a top quark and a $Z$ boson in $pp$ collisions at
  $\sqrt{s} = 13$ TeV with the ATLAS detector''},
\href{http://arxiv.org/abs/2002.07546}{{\ttfamily arXiv:2002.07546 [hep-ex]}}.

\bibitem{Khachatryan:2015ota}
{\bfseries CMS} Collaboration, V.~Khachatryan {\em et~al.}, ``{Search for the
  associated production of a Higgs boson with a single top quark in
  proton-proton collisions at $ \sqrt{s}=8 $ TeV''},
  \href{http://dx.doi.org/10.1007/JHEP06(2016)177}{{\em JHEP} {\bfseries 06}
  (2016) 177},
\href{http://arxiv.org/abs/1509.08159}{{\ttfamily arXiv:1509.08159 [hep-ex]}}.

\bibitem{CMS-PAS-HIG-16-019}
{\bfseries CMS} Collaboration, C.~Collaboration,
``{Search for H to bbar in association with a single top quark as a test of
  Higgs boson couplings at 13 TeV''},.

\bibitem{Sirunyan:2018lzm}
{\bfseries CMS} Collaboration, A.~M. Sirunyan {\em et~al.}, ``{Search for
  associated production of a Higgs boson and a single top quark in
  proton-proton collisions at $\sqrt{s} =$ TeV''},
  \href{http://dx.doi.org/10.1103/PhysRevD.99.092005}{{\em Phys. Rev.}
  {\bfseries D99} no.~9, (2019) 092005},
\href{http://arxiv.org/abs/1811.09696}{{\ttfamily arXiv:1811.09696 [hep-ex]}}.

\bibitem{Maltoni:2001hu}
F.~Maltoni, K.~Paul, T.~Stelzer, and S.~Willenbrock, ``{Associated production
  of Higgs and single top at hadron colliders''},
  \href{http://dx.doi.org/10.1103/PhysRevD.64.094023}{{\em Phys. Rev.}
  {\bfseries D64} (2001) 094023},
\href{http://arxiv.org/abs/hep-ph/0106293}{{\ttfamily arXiv:hep-ph/0106293
  [hep-ph]}}.

\bibitem{Biswas:2012bd}
S.~Biswas, E.~Gabrielli, and B.~Mele, ``{Single top and Higgs associated
  production as a probe of the Htt coupling sign at the LHC''},
  \href{http://dx.doi.org/10.1007/JHEP01(2013)088}{{\em JHEP} {\bfseries 01}
  (2013) 088},
\href{http://arxiv.org/abs/1211.0499}{{\ttfamily arXiv:1211.0499 [hep-ph]}}.

\bibitem{Farina:2012xp}
M.~Farina, C.~Grojean, F.~Maltoni, E.~Salvioni, and A.~Thamm, ``{Lifting
  degeneracies in Higgs couplings using single top production in association
  with a Higgs boson''}, \href{http://dx.doi.org/10.1007/JHEP05(2013)022}{{\em
  JHEP} {\bfseries 05} (2013) 022},
\href{http://arxiv.org/abs/1211.3736}{{\ttfamily arXiv:1211.3736 [hep-ph]}}.

\bibitem{Demartin:2015uha}
F.~Demartin, F.~Maltoni, K.~Mawatari, and M.~Zaro, ``{Higgs production in
  association with a single top quark at the LHC''},
  \href{http://dx.doi.org/10.1140/epjc/s10052-015-3475-9}{{\em Eur. Phys. J.}
  {\bfseries C75} no.~6, (2015) 267},
\href{http://arxiv.org/abs/1504.00611}{{\ttfamily arXiv:1504.00611 [hep-ph]}}.

\bibitem{Degrande:2018fog}
C.~Degrande, F.~Maltoni, K.~Mimasu, E.~Vryonidou, and C.~Zhang, ``{Single-top
  associated production with a $Z$ or $H$ boson at the LHC: the SMEFT
  interpretation''}, \href{http://dx.doi.org/10.1007/JHEP10(2018)005}{{\em
  JHEP} {\bfseries 10} (2018) 005},
\href{http://arxiv.org/abs/1804.07773}{{\ttfamily arXiv:1804.07773 [hep-ph]}}.

\bibitem{Campbell:2013yla}
J.~Campbell, R.~K. Ellis, and R.~Röntsch, ``{Single top production in
  association with a Z boson at the LHC''},
  \href{http://dx.doi.org/10.1103/PhysRevD.87.114006}{{\em Phys. Rev.}
  {\bfseries D87} (2013) 114006},
\href{http://arxiv.org/abs/1302.3856}{{\ttfamily arXiv:1302.3856 [hep-ph]}}.

\bibitem{Amoroso:2020lgh}
S.~Amoroso {\em et~al.}, ``{Les Houches 2019: Physics at TeV Colliders:
  Standard Model Working Group Report''}, in {\em {11th Les Houches Workshop on
  Physics at TeV Colliders: PhysTeV Les Houches (PhysTeV 2019) Les Houches,
  France, June 10-28, 2019}}.
\newblock 2020.
\newblock
\href{http://arxiv.org/abs/2003.01700}{{\ttfamily arXiv:2003.01700 [hep-ph]}}.
\newblock

\bibitem{Alwall:2014hca}
J.~Alwall, R.~Frederix, S.~Frixione, V.~Hirschi, F.~Maltoni, O.~Mattelaer,
  H.~S. Shao, T.~Stelzer, P.~Torrielli, and M.~Zaro, ``{The automated
  computation of tree-level and next-to-leading order differential cross
  sections, and their matching to parton shower simulations''},
  \href{http://dx.doi.org/10.1007/JHEP07(2014)079}{{\em JHEP} {\bfseries 07}
  (2014) 079},
\href{http://arxiv.org/abs/1405.0301}{{\ttfamily arXiv:1405.0301 [hep-ph]}}.

\bibitem{Frederix:2018nkq}
R.~Frederix, S.~Frixione, V.~Hirschi, D.~Pagani, H.~S. Shao, and M.~Zaro,
  ``{The automation of next-to-leading order electroweak calculations''},
  \href{http://dx.doi.org/10.1007/JHEP07(2018)185}{{\em JHEP} {\bfseries 07}
  (2018) 185},
\href{http://arxiv.org/abs/1804.10017}{{\ttfamily arXiv:1804.10017 [hep-ph]}}.

\bibitem{Frixione:2014qaa}
S.~Frixione, V.~Hirschi, D.~Pagani, H.~S. Shao, and M.~Zaro, ``{Weak
  corrections to Higgs hadroproduction in association with a top-quark pair''},
  \href{http://dx.doi.org/10.1007/JHEP09(2014)065}{{\em JHEP} {\bfseries 09}
  (2014) 065},
\href{http://arxiv.org/abs/1407.0823}{{\ttfamily arXiv:1407.0823 [hep-ph]}}.

\bibitem{Frixione:2015zaa}
S.~Frixione, V.~Hirschi, D.~Pagani, H.~S. Shao, and M.~Zaro, ``{Electroweak and
  QCD corrections to top-pair hadroproduction in association with heavy
  bosons''}, \href{http://dx.doi.org/10.1007/JHEP06(2015)184}{{\em JHEP}
  {\bfseries 06} (2015) 184},
\href{http://arxiv.org/abs/1504.03446}{{\ttfamily arXiv:1504.03446 [hep-ph]}}.

\bibitem{Pagani:2016caq}
D.~Pagani, I.~Tsinikos, and M.~Zaro, ``{The impact of the photon PDF and
  electroweak corrections on $t \bar{t}$ distributions''},
  \href{http://dx.doi.org/10.1140/epjc/s10052-016-4318-z}{{\em Eur. Phys. J.}
  {\bfseries C76} no.~9, (2016) 479},
\href{http://arxiv.org/abs/1606.01915}{{\ttfamily arXiv:1606.01915 [hep-ph]}}.

\bibitem{Frederix:2016ost}
R.~Frederix, S.~Frixione, V.~Hirschi, D.~Pagani, H.-S. Shao, and M.~Zaro,
  ``{The complete NLO corrections to dijet hadroproduction''},
  \href{http://dx.doi.org/10.1007/JHEP04(2017)076}{{\em JHEP} {\bfseries 04}
  (2017) 076},
\href{http://arxiv.org/abs/1612.06548}{{\ttfamily arXiv:1612.06548 [hep-ph]}}.

\bibitem{Czakon:2017wor}
M.~Czakon, D.~Heymes, A.~Mitov, D.~Pagani, I.~Tsinikos, and M.~Zaro,
  ``{Top-pair production at the LHC through NNLO QCD and NLO EW''},
  \href{http://dx.doi.org/10.1007/JHEP10(2017)186}{{\em JHEP} {\bfseries 10}
  (2017) 186},
\href{http://arxiv.org/abs/1705.04105}{{\ttfamily arXiv:1705.04105 [hep-ph]}}.

\bibitem{Frederix:2017wme}
R.~Frederix, D.~Pagani, and M.~Zaro, ``{Large NLO corrections in
  $t\bar{t}W^{\pm}$ and $t\bar{t}t\bar{t}$ hadroproduction from supposedly
  subleading EW contributions''},
  \href{http://dx.doi.org/10.1007/JHEP02(2018)031}{{\em JHEP} {\bfseries 02}
  (2018) 031},
\href{http://arxiv.org/abs/1711.02116}{{\ttfamily arXiv:1711.02116 [hep-ph]}}.

\bibitem{Broggio:2019ewu}
A.~Broggio, A.~Ferroglia, R.~Frederix, D.~Pagani, B.~D. Pecjak, and
  I.~Tsinikos, ``{Top-quark pair hadroproduction in association with a heavy
  boson at NLO+NNLL including EW corrections''},
  \href{http://dx.doi.org/10.1007/JHEP08(2019)039}{{\em JHEP} {\bfseries 08}
  (2019) 039},
\href{http://arxiv.org/abs/1907.04343}{{\ttfamily arXiv:1907.04343 [hep-ph]}}.

\bibitem{Frederix:2019ubd}
R.~Frederix, D.~Pagani, and I.~Tsinikos, ``{Precise predictions for single-top
  production: the impact of EW corrections and QCD shower on the $t$-channel
  signature''}, \href{http://dx.doi.org/10.1007/JHEP09(2019)122}{{\em JHEP}
  {\bfseries 09} (2019) 122},
\href{http://arxiv.org/abs/1907.12586}{{\ttfamily arXiv:1907.12586 [hep-ph]}}.

\bibitem{Kulesza:2020nfh}
A.~Kulesza, L.~Motyka, D.~Schwartländer, T.~Stebel, and V.~Theeuwes,
  ``{Associated top quark pair production with a heavy boson: differential
  cross sections at NLO+NNLL accuracy''},
\href{http://arxiv.org/abs/2001.03031}{{\ttfamily arXiv:2001.03031 [hep-ph]}}.

\bibitem{Frederix:2020jzp}
R.~Frederix and I.~Tsinikos, ``{Subleading EW corrections and spin-correlation
  effects in $t\bar{t}W$ multi-lepton signatures''},
  \href{http://arxiv.org/abs/2004.09552}{{\ttfamily arXiv:2004.09552
  [hep-ph]}}.


\bibitem{Maltoni:2012pa}
F.~Maltoni, G.~Ridolfi, and M.~Ubiali, ``{b-initiated processes at the LHC: a
  reappraisal''}, \href{http://dx.doi.org/10.1007/JHEP04(2013)095,
  10.1007/JHEP07(2012)022}{{\em JHEP} {\bfseries 07} (2012) 022},
  \href{http://arxiv.org/abs/1203.6393}{{\ttfamily arXiv:1203.6393 [hep-ph]}}.
[Erratum: JHEP04,095(2013)].

\bibitem{Lim:2016wjo}
M.~Lim, F.~Maltoni, G.~Ridolfi, and M.~Ubiali, ``{Anatomy of double heavy-quark
  initiated processes''}, \href{http://dx.doi.org/10.1007/JHEP09(2016)132}{{\em
  JHEP} {\bfseries 09} (2016) 132},
\href{http://arxiv.org/abs/1605.09411}{{\ttfamily arXiv:1605.09411 [hep-ph]}}.

\bibitem{Frederix:2013gra}
R.~Frederix, ``{Top Quark Induced Backgrounds to Higgs Production in the
  $WW^{(*)}\to ll\nu\nu$ Decay Channel at Next-to-Leading-Order in QCD''},
  \href{http://dx.doi.org/10.1103/PhysRevLett.112.082002}{{\em Phys. Rev.
  Lett.} {\bfseries 112} no.~8, (2014) 082002},
\href{http://arxiv.org/abs/1311.4893}{{\ttfamily arXiv:1311.4893 [hep-ph]}}.

\bibitem{Cascioli:2013wga}
F.~Cascioli, S.~Kallweit, P.~Maierhöfer, and S.~Pozzorini, ``{A unified NLO
  description of top-pair and associated Wt production''},
  \href{http://dx.doi.org/10.1140/epjc/s10052-014-2783-9}{{\em Eur. Phys. J.}
  {\bfseries C74} no.~3, (2014) 2783},
\href{http://arxiv.org/abs/1312.0546}{{\ttfamily arXiv:1312.0546 [hep-ph]}}.

\bibitem{Jezo:2016ujg}
T.~Je\v{z}o, J.~M. Lindert, P.~Nason, C.~Oleari, and S.~Pozzorini, ``{An NLO+PS
  generator for $t\bar{t}$ and $Wt$ production and decay including non-resonant
  and interference effects''},
  \href{http://dx.doi.org/10.1140/epjc/s10052-016-4538-2}{{\em Eur. Phys. J.}
  {\bfseries C76} no.~12, (2016) 691},
\href{http://arxiv.org/abs/1607.04538}{{\ttfamily arXiv:1607.04538 [hep-ph]}}.

\bibitem{Frixione:2008yi}
S.~Frixione, E.~Laenen, P.~Motylinski, B.~R. Webber, and C.~D. White,
  ``{Single-top hadroproduction in association with a W boson''},
  \href{http://dx.doi.org/10.1088/1126-6708/2008/07/029}{{\em JHEP} {\bfseries
  07} (2008) 029},
\href{http://arxiv.org/abs/0805.3067}{{\ttfamily arXiv:0805.3067 [hep-ph]}}.

\bibitem{Hollik:2012rc}
W.~Hollik, J.~M. Lindert, and D.~Pagani, ``{NLO corrections to squark-squark
  production and decay at the LHC''},
  \href{http://dx.doi.org/10.1007/JHEP03(2013)139}{{\em JHEP} {\bfseries 03}
  (2013) 139},
\href{http://arxiv.org/abs/1207.1071}{{\ttfamily arXiv:1207.1071 [hep-ph]}}.

\bibitem{Beenakker:1996ch}
W.~Beenakker, R.~Hopker, M.~Spira, and P.~M. Zerwas, ``{Squark and gluino
  production at hadron colliders''},
  \href{http://dx.doi.org/10.1016/S0550-3213(97)80027-2}{{\em Nucl. Phys.}
  {\bfseries B492} (1997) 51--103},
\href{http://arxiv.org/abs/hep-ph/9610490}{{\ttfamily arXiv:hep-ph/9610490
  [hep-ph]}}.

\bibitem{Gavin:2013kga}
R.~Gavin, C.~Hangst, M.~Krämer, M.~Mühlleitner, M.~Pellen, E.~Popenda, and
  M.~Spira, ``{Matching Squark Pair Production at NLO with Parton Showers''},
  \href{http://dx.doi.org/10.1007/JHEP10(2013)187}{{\em JHEP} {\bfseries 10}
  (2013) 187},
\href{http://arxiv.org/abs/1305.4061}{{\ttfamily arXiv:1305.4061 [hep-ph]}}.

\bibitem{Demartin:2016axk}
F.~Demartin, B.~Maier, F.~Maltoni, K.~Mawatari, and M.~Zaro, ``{tWH associated
  production at the LHC''},
  \href{http://dx.doi.org/10.1140/epjc/s10052-017-4601-7}{{\em Eur. Phys. J.}
  {\bfseries C77} no.~1, (2017) 34},
\href{http://arxiv.org/abs/1607.05862}{{\ttfamily arXiv:1607.05862 [hep-ph]}}.

\bibitem{Pagani:2020rsg}
D.~Pagani, H.-S. Shao, and M.~Zaro, ``{RIP $H b \bar b$: How other Higgs
  production modes conspire to kill a rare signal at the LHC''},
  \href{http://arxiv.org/abs/2005.10277}{{\ttfamily arXiv:2005.10277
  [hep-ph]}}.

\bibitem{Denner:1999gp}
A.~Denner, S.~Dittmaier, M.~Roth, and D.~Wackeroth, ``{Predictions for all
  processes e+ e- ---> 4 fermions + gamma''},
  \href{http://dx.doi.org/10.1016/S0550-3213(99)00437-X}{{\em Nucl. Phys. B}
  {\bfseries 560} (1999) 33--65},
  \href{http://arxiv.org/abs/hep-ph/9904472}{{\ttfamily arXiv:hep-ph/9904472}}.

\bibitem{Denner:2005fg}
A.~Denner, S.~Dittmaier, M.~Roth, and L.~Wieders, ``{Electroweak corrections to
  charged-current e+ e- ---> 4 fermion processes: Technical details and further
  results''}, \href{http://dx.doi.org/10.1016/j.nuclphysb.2011.09.001}{{\em
  Nucl. Phys. B} {\bfseries 724} (2005) 247--294},
  \href{http://arxiv.org/abs/hep-ph/0505042}{{\ttfamily arXiv:hep-ph/0505042}}.
  [Erratum: Nucl.Phys.B 854, 504--507 (2012)].

\bibitem{Ball:2014uwa}
{\bfseries NNPDF} Collaboration, R.~D. Ball {\em et~al.}, ``{Parton
  distributions for the LHC Run II''},
  \href{http://dx.doi.org/10.1007/JHEP04(2015)040}{{\em JHEP} {\bfseries 04}
  (2015) 040},
\href{http://arxiv.org/abs/1410.8849}{{\ttfamily arXiv:1410.8849 [hep-ph]}}.

\bibitem{Bertone:2016ume}
V.~Bertone and S.~Carrazza, ``{Combining NNPDF3.0 and NNPDF2.3QED through the
  APFEL evolution code''}, \href{http://dx.doi.org/10.22323/1.265.0031}{{\em
  PoS} {\bfseries DIS2016} (2016) 031},
\href{http://arxiv.org/abs/1606.07130}{{\ttfamily arXiv:1606.07130 [hep-ph]}}.

\bibitem{Ball:2017nwa}
{\bfseries NNPDF} Collaboration, R.~D. Ball {\em et~al.}, ``{Parton
  distributions from high-precision collider data''},
  \href{http://dx.doi.org/10.1140/epjc/s10052-017-5199-5}{{\em Eur. Phys. J.}
  {\bfseries C77} no.~10, (2017) 663},
\href{http://arxiv.org/abs/1706.00428}{{\ttfamily arXiv:1706.00428 [hep-ph]}}.

\bibitem{Bertone:2017bme}
{\bfseries NNPDF} Collaboration, V.~Bertone, S.~Carrazza, N.~P. Hartland, and
  J.~Rojo, ``{Illuminating the photon content of the proton within a global PDF
  analysis''}, \href{http://dx.doi.org/10.21468/SciPostPhys.5.1.008}{{\em
  SciPost Phys.} {\bfseries 5} no.~1, (2018) 008},
\href{http://arxiv.org/abs/1712.07053}{{\ttfamily arXiv:1712.07053 [hep-ph]}}.

\bibitem{Harland-Lang:2014zoa}
L.~A. Harland-Lang, A.~D. Martin, P.~Motylinski, and R.~S. Thorne, ``{Parton
  distributions in the LHC era: MMHT 2014 PDFs''},
  \href{http://dx.doi.org/10.1140/epjc/s10052-015-3397-6}{{\em Eur. Phys. J.}
  {\bfseries C75} no.~5, (2015) 204},
\href{http://arxiv.org/abs/1412.3989}{{\ttfamily arXiv:1412.3989 [hep-ph]}}.

\bibitem{Harland-Lang:2019pla}
L.~A. Harland-Lang, A.~D. Martin, R.~Nathvani, and R.~S. Thorne, ``{Ad Lucem:
  QED Parton Distribution Functions in the MMHT Framework''},
  \href{http://dx.doi.org/10.1140/epjc/s10052-019-7296-0}{{\em Eur. Phys. J.}
  {\bfseries C79} no.~10, (2019) 811},
\href{http://arxiv.org/abs/1907.02750}{{\ttfamily arXiv:1907.02750 [hep-ph]}}.

\bibitem{Manohar:2016nzj}
A.~Manohar, P.~Nason, G.~P. Salam, and G.~Zanderighi, ``{How bright is the
  proton? A precise determination of the photon parton distribution
  function''}, \href{http://dx.doi.org/10.1103/PhysRevLett.117.242002}{{\em
  Phys. Rev. Lett.} {\bfseries 117} no.~24, (2016) 242002},
\href{http://arxiv.org/abs/1607.04266}{{\ttfamily arXiv:1607.04266 [hep-ph]}}.

\bibitem{Manohar:2017eqh}
A.~V. Manohar, P.~Nason, G.~P. Salam, and G.~Zanderighi, ``{The Photon Content
  of the Proton''}, \href{http://dx.doi.org/10.1007/JHEP12(2017)046}{{\em JHEP}
  {\bfseries 12} (2017) 046},
\href{http://arxiv.org/abs/1708.01256}{{\ttfamily arXiv:1708.01256 [hep-ph]}}.

\bibitem{Ball:2011mu}
R.~D. Ball, V.~Bertone, F.~Cerutti, L.~Del~Debbio, S.~Forte, A.~Guffanti, J.~I.
  Latorre, J.~Rojo, and M.~Ubiali, ``{Impact of Heavy Quark Masses on Parton
  Distributions and LHC Phenomenology''},
  \href{http://dx.doi.org/10.1016/j.nuclphysb.2011.03.021}{{\em Nucl. Phys.}
  {\bfseries B849} (2011) 296--363},
\href{http://arxiv.org/abs/1101.1300}{{\ttfamily arXiv:1101.1300 [hep-ph]}}.

\bibitem{Cacciari:2008gp}
M.~Cacciari, G.~P. Salam, and G.~Soyez, ``{The anti-$k_t$ jet clustering
  algorithm''}, \href{http://dx.doi.org/10.1088/1126-6708/2008/04/063}{{\em
  JHEP} {\bfseries 04} (2008) 063},
\href{http://arxiv.org/abs/0802.1189}{{\ttfamily arXiv:0802.1189 [hep-ph]}}.

\bibitem{Cacciari:2011ma}
M.~Cacciari, G.~P. Salam, and G.~Soyez, ``{FastJet User Manual''},
  \href{http://dx.doi.org/10.1140/epjc/s10052-012-1896-2}{{\em Eur. Phys. J. C}
  {\bfseries 72} (2012) 1896}, \href{http://arxiv.org/abs/1111.6097}{{\ttfamily
  arXiv:1111.6097 [hep-ph]}}.


\bibitem{Frixione:1995ms}
S.~Frixione, Z.~Kunszt, and A.~Signer, ``{Three jet cross-sections to
  next-to-leading order''},
  \href{http://dx.doi.org/10.1016/0550-3213(96)00110-1}{{\em Nucl. Phys.}
  {\bfseries B467} (1996) 399--442},
\href{http://arxiv.org/abs/hep-ph/9512328}{{\ttfamily arXiv:hep-ph/9512328
  [hep-ph]}}.

\bibitem{Frixione:1997np}
S.~Frixione, ``{A General approach to jet cross-sections in QCD''},
  \href{http://dx.doi.org/10.1016/S0550-3213(97)00574-9}{{\em Nucl. Phys.}
  {\bfseries B507} (1997) 295--314},
\href{http://arxiv.org/abs/hep-ph/9706545}{{\ttfamily arXiv:hep-ph/9706545
  [hep-ph]}}.

\bibitem{Frederix:2009yq}
R.~Frederix, S.~Frixione, F.~Maltoni, and T.~Stelzer, ``{Automation of
  next-to-leading order computations in QCD: The FKS subtraction''},
  \href{http://dx.doi.org/10.1088/1126-6708/2009/10/003}{{\em JHEP} {\bfseries
  10} (2009) 003},
\href{http://arxiv.org/abs/0908.4272}{{\ttfamily arXiv:0908.4272 [hep-ph]}}.

\bibitem{Frederix:2016rdc}
R.~Frederix, S.~Frixione, A.~S. Papanastasiou, S.~Prestel, and P.~Torrielli,
  ``{Off-shell single-top production at NLO matched to parton showers''},
  \href{http://dx.doi.org/10.1007/JHEP06(2016)027}{{\em JHEP} {\bfseries 06}
  (2016) 027},
\href{http://arxiv.org/abs/1603.01178}{{\ttfamily arXiv:1603.01178 [hep-ph]}}.

\bibitem{Ossola:2006us}
G.~Ossola, C.~G. Papadopoulos, and R.~Pittau, ``{Reducing full one-loop
  amplitudes to scalar integrals at the integrand level''},
  \href{http://dx.doi.org/10.1016/j.nuclphysb.2006.11.012}{{\em Nucl. Phys.}
  {\bfseries B763} (2007) 147--169},
\href{http://arxiv.org/abs/hep-ph/0609007}{{\ttfamily arXiv:hep-ph/0609007
  [hep-ph]}}.

\bibitem{Mastrolia:2012bu}
P.~Mastrolia, E.~Mirabella, and T.~Peraro, ``{Integrand reduction of one-loop
  scattering amplitudes through Laurent series expansion''},
  \href{http://dx.doi.org/10.1007/JHEP11(2012)128,
  10.1007/JHEP06(2012)095}{{\em JHEP} {\bfseries 06} (2012) 095},
  \href{http://arxiv.org/abs/1203.0291}{{\ttfamily arXiv:1203.0291 [hep-ph]}}.
[Erratum: JHEP11,128(2012)].

\bibitem{Passarino:1978jh}
G.~Passarino and M.~J.~G. Veltman, ``{One Loop Corrections for e+ e-
  Annihilation Into mu+ mu- in the Weinberg Model''},
\href{http://dx.doi.org/10.1016/0550-3213(79)90234-7}{{\em Nucl. Phys.}
  {\bfseries B160} (1979) 151--207}.

\bibitem{Davydychev:1991va}
A.~I. Davydychev, ``{A Simple formula for reducing Feynman diagrams to scalar
  integrals''},
\href{http://dx.doi.org/10.1016/0370-2693(91)91715-8}{{\em Phys. Lett.}
  {\bfseries B263} (1991) 107--111}.

\bibitem{Denner:2005nn}
A.~Denner and S.~Dittmaier, ``{Reduction schemes for one-loop tensor
  integrals''}, \href{http://dx.doi.org/10.1016/j.nuclphysb.2005.11.007}{{\em
  Nucl. Phys.} {\bfseries B734} (2006) 62--115},
\href{http://arxiv.org/abs/hep-ph/0509141}{{\ttfamily arXiv:hep-ph/0509141
  [hep-ph]}}.

\bibitem{Hirschi:2011pa}
V.~Hirschi, R.~Frederix, S.~Frixione, M.~V. Garzelli, F.~Maltoni, and
  R.~Pittau, ``{Automation of one-loop QCD corrections''},
  \href{http://dx.doi.org/10.1007/JHEP05(2011)044}{{\em JHEP} {\bfseries 05}
  (2011) 044},
\href{http://arxiv.org/abs/1103.0621}{{\ttfamily arXiv:1103.0621 [hep-ph]}}.

\bibitem{Ossola:2007ax}
G.~Ossola, C.~G. Papadopoulos, and R.~Pittau, ``{CutTools: A Program
  implementing the OPP reduction method to compute one-loop amplitudes''},
  \href{http://dx.doi.org/10.1088/1126-6708/2008/03/042}{{\em JHEP} {\bfseries
  03} (2008) 042},
\href{http://arxiv.org/abs/0711.3596}{{\ttfamily arXiv:0711.3596 [hep-ph]}}.

\bibitem{Peraro:2014cba}
T.~Peraro, ``{Ninja: Automated Integrand Reduction via Laurent Expansion for
  One-Loop Amplitudes''},
  \href{http://dx.doi.org/10.1016/j.cpc.2014.06.017}{{\em Comput. Phys.
  Commun.} {\bfseries 185} (2014) 2771--2797},
\href{http://arxiv.org/abs/1403.1229}{{\ttfamily arXiv:1403.1229 [hep-ph]}}.

\bibitem{Hirschi:2016mdz}
V.~Hirschi and T.~Peraro, ``{Tensor integrand reduction via Laurent
  expansion''}, \href{http://dx.doi.org/10.1007/JHEP06(2016)060}{{\em JHEP}
  {\bfseries 06} (2016) 060},
\href{http://arxiv.org/abs/1604.01363}{{\ttfamily arXiv:1604.01363 [hep-ph]}}.

\bibitem{Denner:2016kdg}
A.~Denner, S.~Dittmaier, and L.~Hofer, ``{Collier: a fortran-based Complex
  One-Loop LIbrary in Extended Regularizations''},
  \href{http://dx.doi.org/10.1016/j.cpc.2016.10.013}{{\em Comput. Phys.
  Commun.} {\bfseries 212} (2017) 220--238},
\href{http://arxiv.org/abs/1604.06792}{{\ttfamily arXiv:1604.06792 [hep-ph]}}.

\bibitem{Cascioli:2011va}
F.~Cascioli, P.~Maierhofer, and S.~Pozzorini, ``{Scattering Amplitudes with
  Open Loops''}, \href{http://dx.doi.org/10.1103/PhysRevLett.108.111601}{{\em
  Phys. Rev. Lett.} {\bfseries 108} (2012) 111601},
\href{http://arxiv.org/abs/1111.5206}{{\ttfamily arXiv:1111.5206 [hep-ph]}}.




\bibitem{Campbell:2009ss}
J.~M. Campbell, R.~Frederix, F.~Maltoni, and F.~Tramontano,
  ``{Next-to-Leading-Order Predictions for t-Channel Single-Top Production at
  Hadron Colliders''},
  \href{http://dx.doi.org/10.1103/PhysRevLett.102.182003}{{\em Phys. Rev.
  Lett.} {\bfseries 102} (2009) 182003},
\href{http://arxiv.org/abs/0903.0005}{{\ttfamily arXiv:0903.0005 [hep-ph]}}.

\bibitem{Sjostrand:2007gs}
T.~Sjöstrand, S.~Mrenna, and P.~Z. Skands, ``{A Brief Introduction to PYTHIA
  8.1''}, \href{http://dx.doi.org/10.1016/j.cpc.2008.01.036}{{\em Comput. Phys.
  Commun.} {\bfseries 178} (2008) 852--867},
  \href{http://arxiv.org/abs/0710.3820}{{\ttfamily arXiv:0710.3820 [hep-ph]}}.

\bibitem{Sjostrand:2014zea}
T.~Sjöstrand, S.~Ask, J.~R. Christiansen, R.~Corke, N.~Desai, P.~Ilten,
  S.~Mrenna, S.~Prestel, C.~O. Rasmussen, and P.~Z. Skands, ``{An Introduction
  to PYTHIA 8.2''}, \href{http://dx.doi.org/10.1016/j.cpc.2015.01.024}{{\em
  Comput. Phys. Commun.} {\bfseries 191} (2015) 159--177},
  \href{http://arxiv.org/abs/1410.3012}{{\ttfamily arXiv:1410.3012 [hep-ph]}}.

\bibitem{Frixione:2002ik}
S.~Frixione and B.~R. Webber, ``{Matching NLO QCD computations and parton
  shower simulations''},
  \href{http://dx.doi.org/10.1088/1126-6708/2002/06/029}{{\em JHEP} {\bfseries
  06} (2002) 029}, \href{http://arxiv.org/abs/hep-ph/0204244}{{\ttfamily
  arXiv:hep-ph/0204244}}.

\end{thebibliography}\endgroup
 
\end{document}